\documentclass[a4paper,11pt]{article}
\pdfoutput=1
\usepackage{appendix}
\usepackage{jcappub}
\usepackage[T1]{fontenc}
\usepackage{aas_macros}

\title{Super sample covariance and the volume scaling of galaxy survey covariance matrices}

\author[a,b]{Greg Schreiner}
\author[a,b,*]{Alex Krolewski}
\author[c,d,a,b]{Shahab Joudaki}
\author[a,b,e]{Will J. Percival}

\affiliation[a]{Waterloo Centre for Astrophysics, University of Waterloo, 200 University Ave W, Waterloo, ON N2L 3G1, Canada}
\affiliation[b]{Department of Physics and Astronomy, University of Waterloo, 200 University Ave W, Waterloo, ON N2L 3G1, Canada}
\affiliation[c]{Centro de Investigaciones Energ\'{e}ticas, Medioambientales y Tecnol\'{o}gicas (CIEMAT), Avenida Complutense 40, E-28040 Madrid, Spain}
\affiliation[d]{Institute of Cosmology and Gravitation, University of Portsmouth, Burnaby Road, Portsmouth, PO1 3FX, United Kingdom}
\affiliation[e]{Perimeter Institute for Theoretical Physics, 31 Caroline St. North, Waterloo, ON N2L 2Y5, Canada}
\affiliation[*]{CITA National Fellow}

\emailAdd{gjschreiner@uwaterloo.ca}

\abstract{Super sample covariance (SSC) is important when estimating covariance matrices using a set of mock catalogues for galaxy surveys. If the underlying cosmological simulations do not include the variation in background parameters appropriate for the simulation sizes, then the scatter between mocks will be missing the SSC component. The coupling between large and small modes due to non-linear structure growth makes this pernicious on small scales. We compare different methods for generating ensembles of mocks with SSC built in to the covariance, and contrast against methods where the SSC component is computed and added to the covariance separately. We find that several perturbative expansions, developed to derive background fluctuations, give similar results. We then consider scaling covariance matrices calculated for simulations of different volumes to improve the accuracy of covariance matrix estimation for a given amount of computational time. On large scales, we find that the primary limitation is from the discrete number of modes contributing to the measured power spectrum, and we propose a new method for correcting this effect. Correct implementation of SSC and the effect of discrete mode numbers allows covariance matrices created from mocks to be scaled between volumes, potentially leading to a significant saving on computational resources when producing covariance matrices. We argue that a sub-percent match is difficult to achieve because of the effects of modes on scales between the box sizes, which cannot be easily included. Even so, when working in real space and cubic boxes, we show that a 3$\%$ match in the dark matter power spectrum covariance is achievable on scales of interest for current surveys scaling the simulation volume by $512\times$, costing a small fraction of the computational time of running full-sized simulations. This is comparable to the agreement between analytic and mock-based covariance estimates to be used with DESI Y1 results.}
\begin{document}
\maketitle
\flushbottom

\section{Introduction} \label{Sec_Introduction}

The next generation of cosmological surveys (DESI \cite{DESI_2016}, Euclid \cite{Euclid_2011}, LSST \cite{LSST}, Roman \cite{roman}, SphereX \cite{spherex}) are set to measure cosmic structure over larger volumes and at higher redshift than ever before. Inferring cosmological parameters from these surveys typically proceeds with a measurement of both an observable (e.g.~the power spectrum) and an estimate of its covariance. In the linear regime, the density field is a Gaussian random field and all statistical information is contained in the power spectrum. These next-generation surveys are pushing deeper into the non-linear regime; since the assumption of Gaussianity is expected to break down in this regime, exploring alternative statistics (e.g.~the bispectrum) is an active area of investigation.

The information of current and near-future analyses is still predominantly contained in the power spectrum \cite{DESI_2016,BOSS}, and hereafter we will consider power spectrum-based analyses. The power spectrum covariance is a 4-point function of the density field and can be estimated in three ways: analytically with a number of approximations, from large ensembles of mock galaxy catalogues, or from the data itself using jackknife type methods. Covariance estimation is often the dominant computational cost for large-scale structure analysis and will continue to rise as surveys push to larger volumes and smaller scales.

The covariance matrix depends on cosmology; strictly, each cosmological model being tested requires a new covariance matrix in the likelihood as we wish to determine the probability of the data given the model \cite{White_2015,Kalus_2016}. Because of the computational expense of this, typically, a single cosmology is used to calculate the covariance used for all models, with parameters similar to the best-fit to the survey data. This is not as bad an approximation as it sounds as the variation of the likelihood caused by the fixed-covariance is similar and opposite to the change from assuming a Gaussian likelihood \citep{Wang2019}, another popular approximation. The assumption of a Gaussian likelihood and fixed covariance matrix was adopted in the recent Dark Energy Spectroscopic Instrument (DESI) results, for example \cite{desi-y1-bao}. The Gaussian likelihood is also well-justified by the central limit theorem for power spectrum bins containing a sufficiently large number of independent modes. In the following, we therefore focus on the calculation of a covariance matrix, calculated at a single model cosmology.

In the linear regime, the covariance matrix can be calculated analytically from the power spectrum \cite{FKP1993} and shot-noise. Beyond the linear regime, the covariance becomes a complicated function dependent on the survey window function, non-Poissonian shot noise, scale-dependent galaxy bias, and small-scale redshift space distortions. Great progress has been made in modeling the dark matter power spectrum covariance \cite{Bertolini_2016,Carron_2015,Hou_2022,Mohammed_2014,Neyrinck_2011,Seljak_2000,Taylor_2013} and the galaxy power spectrum covariance \cite{Sugiyama_2020,Wadekar20} accurately into the non-linear regime, often using perturbation theory or effective field theory of large-scale structure \cite{Bertolini_2016,Wadekar20}.
These analytical methods have the advantage of being less computationally expensive than numerical methods, but much work remains to correctly model nonlinear evolution. Alternatively, one can estimate the covariance matrix from an ensemble of $N$-body simulations. While $N$-body simulations are better able to capture the effects of small-scale evolution than existing analytic models, the simulations are highly computationally expensive to run; each must be large enough to cover the entire survey volume and have high enough resolution to reproduce the observed galaxy distribution. This computational cost can be offset by use of control variates \cite{Chartier_2021,Chartier_2022}, combining the results from a small number of high accuracy simulations with those from a large number of lower accuracy surrogate simulations. While this technique has already demonstrated an order of magnitude or more improvement on covariance matrix precision for a given computation time, the limiting factor is the time it takes to run the high accuracy simulations.

In order not to degrade survey results, the covariance matrix must be modeled with a high enough precision that error in the covariance matrix is sub-dominant to the statistical error from the survey parameters. For next-generation surveys, this will require ensembles of $\mathcal{O}(10,000)$ mocks \cite{Dodelson_2013,Percival_2014,Taylor_2013} if the mocks are used to calculate all covariance matrix elements independently. 

With galaxy survey volume and resolution growing faster than the availability of computational resources for covariance estimation, many studies have aimed to reduce the number of simulations needed to achieve the desired covariance precision. Paz and Sanchez \cite{Paz_2015} presented the covariance tapering method in which bins of the covariance matrix with low signal-to-noise can be down-weighted to reduce noise on measured parameters. Shrinkage can be used to balance an accurate but imprecise covariance made from high-resolution simulations to a precise but inaccurate covariance made either from low-resolution simulations or analytic methods \cite{Pope2008,Joachimi2017}. Alternatively, modeling can be used to reduce the free parameters fitted by the mocks \cite{Gaztanaga2005,Friedrich2018,Pearson2016,Fumagalli2024}. A more extreme option is to use data compression algorithms such as MOPED \cite{MOPED} to drastically reduce the size of the data \cite{Lai-2024}, and hence the number of mocks required to calculate a covariance matrix.
Rather than reducing the number of mocks required, \cite{Howlett_2017} demonstrated a method where the covariance matrix can be estimated using an ensemble of smaller mocks, leveraging the volume scaling of the covariance. 

One complication with any mock-based approach is the lack of modes in simulations on scales larger than the box size, which affect the measured covariance on both large and small scales. This effect is called super-sample covariance (SSC). For the large-scale covariance to be recovered without correction, the box must be large enough to include all modes that affect the covariance within a survey. The substantial coupling between large-scale and small-scale modes makes this difficult to achieve. The UCHUU-GLAM project \cite{Klypin_2018,Klypin_2019,DongPaez24,Ereza_2024} has shown great success in modeling the covariance of the BOSS and eBOSS luminous red galaxy samples by re-scaling results from simulations smaller than the samples but large enough that the SSC contribution is insignificant ($L\geq 1.5\ h^{-1}$ Gpc). To scale from even smaller (and thereby cheaper) mocks, the SSC effect must be properly accounted for to recover an accurate estimate of the covariance from small mocks.

In this work, we evaluate different prescriptions for including super-sample covariance in an ensemble of simulations. We estimate the covariance matrix from different ensembles of simulations and compare them to the covariance matrix from subsampled mocks to determine how well each prescription recovers the true SSC effect. We then use the volume scaling technique to recover large mock covariance from ensembles of smaller volume mocks and determine conditions for which the volume scaling of the covariance breaks down. This establishes the limit in which the volume scaling technique can be leveraged to maximally reduce computation time without significantly biasing the covariance matrix estimate. For simplicity we focus on the dark matter power spectrum, without a window function, and ignore redshift-space distortions. We do not expect these simplifications to significantly affect our general conclusions about the validity of the approach. 

The layout of our paper is as follows. In Section~\ref{Sec_CovEstimation}, we outline the origin of the SSC effect as a power spectrum response to a background mode and review different methods of including SSC in simulations. In Section~\ref{Sec_CompareSSCModels}, we evaluate the effectiveness of these different methods in recovering the SSC effect. In Section~\ref{Sec_VolumeScaling}, we test the volume scaling behaviour of the covariance matrix and determine where it breaks down. In Section~\ref{Sec_Conclusions}, we discuss these findings and identify the conditions in which volume scaling provides the greatest increase in computational efficiency without loss of information in the covariance matrix at scales of interest.

\section{Covariance estimation from simulations} \label{Sec_CovEstimation}
In this section, we describe how the covariance matrix is most simply estimated using $N$-body simulations, the effect of super-sample covariance, and how this super-sample effect can be captured by running separate universe simulations.
\subsection{The power spectrum and its covariance} \label{SS_PSCov}
The distribution of matter in a survey or simulation volume is often represented by the overdensity field $\delta (\mathbf{x})$. The most straightforward way to measure the clustering of matter is using the matter power spectrum, estimated by
\begin{equation}
    P(\mathbf{k}_i) = \left< \tilde{\delta}(\mathbf{k}_i) \tilde{\delta}^*(\mathbf{k}_i) \right>\,,
\end{equation}
where $\tilde{\delta}(\mathbf{k})$ is the Fourier transform of the overdensity field. The statistical uncertainty of this estimator is captured by its covariance matrix 
\begin{equation}
    C(k_i,k_j) = \left< P(k_i) P(k_j)\right> - \left< P(k_i) \right> \left< P(k_j) \right>\,,
    \label{eqn_Covariance}
\end{equation}
where $\left< \right>$ denotes the average value over the ensemble of realizations. The covariance matrix is a 4-point clustering statistic. In the case of $\delta$ being a Gaussian random field in a cubic survey volume with periodic boundary conditions, this means that the covariance matrix can be calculated directly from the power spectrum
\begin{equation}
    C_{\textrm{Gauss}}(k_i,k_j) \equiv \frac{1}{V}\frac{(2\pi)^3}{V_{k_i}}2P(k_i)^2\delta^K_{ij}\,,
    \label{eqn_Cgauss}
\end{equation}
where $V$ is the survey volume, $V_{k_i}$ is the volume of the $i^{th}$ spherical shell in $k$ space, and $\delta^K_{ij}$ is the Kronecker delta function. In this case the covariance $C_{\textrm{Gauss}}$ would be diagonal, reflecting the fact that the modes of a Gaussian random field are independent of each other. Once nonlinear evolution has occurred and modes of the power spectrum become correlated, the off diagonal terms of the covariance become nonzero. In this case, it has been established that the covariance matrix picks up a contribution from the matter trispectrum \cite{FKP1993,Meiksin_1999,Scoccimarro_1999,Takada_2013}
\begin{equation}
    C(k_i,k_j) = C_{\textrm{Gauss}}(k_i,k_j) + \frac{1}{V}\overline{T}(k_i,k_j), \label{eqn_CnoSSC}
\end{equation}
were $\overline{T}$ is the bin-averaged trispectrum
\begin{equation}
    \overline{T}(k_i,k_j) = \int_{|\mathbf{k}|\in k_i}\frac{d^3\mathbf{k}}{V_{k_i}} \int_{|\mathbf{k}'|\in k_j}\frac{d^3\mathbf{k}'}{V_{k_j}} T(\mathbf{k},-\mathbf{k},\mathbf{k}',-\mathbf{k}')\,. \label{eqn_trispectrum}
\end{equation}
Given an ensemble of realizations of the density field, a brute-force estimate of the covariance matrix can be constructed
\begin{equation}
    C(k_i,k_j) = \frac{1}{N_s - 1}\sum_{m=1}^{N_s} \left[ P_m(k_i) - \overline{P}(k_i) \right] \left[ P_m(k_j) - \overline{P}(k_j) \right]\,,
    \label{eqn_covariance_estimate}
\end{equation}
where $P_m(k_i)$ is the $i^{th}$ bin of the $m^{th}$ power spectrum sample, and $\overline{P}(k_i)$ is the average power among all $N_s$ samples in the $i^{th}$ bin.

\subsection{Super-sample covariance}
The long- and short- wavelength modes of the power spectrum are coupled due to non-linear gravitational evolution. Consequently, when measuring the power in a survey volume, modes larger than the survey volume have a significant effect on the modes within the survey despite not being able to be measured directly. The variance of these super-survey modes thus contributes to the covariance matrix on scales of interest within the survey. This additional covariance is termed super-sample covariance.

Super-sample covariance has been the subject of many studies given its importance in determining errors \cite{Hu_2003,Hamilton_2006,Gnedin_2011,Takahashi_2009,Baldauf_2011,Takada_2013,Mohammed_2014,Li_2014,Wagner_2015,Akitsu_2017,Akitsu_2018,Barreira_2017a,Barreira_2017b,Howlett_2017,Chan_2018,Li_2018,Barreira_2018,Barreira_2019a,Barreira_2019b,Lacasa_2019,Castorina_2020,Philcox_2020,Halder_2022,Zhai_2022,Bayer_2022,Gouyou_2022,Terasawa_2022,Linke_2024}. Extracting information from the weakly non-linear and fully non-linear scales of the power spectrum requires accurate modeling of super-sample covariance due to its significant contribution to the covariance in these regimes.

In the work of Ref.~\cite{Takada_2013}, the predominant effect of a super-sample mode can be conceptualized as a rescaling of the background density $\delta_b$ of the survey or simulation. To measure super-sample covariance from simulations, one must determine the response of the power spectrum to a change in $\delta_b$. The covariance can then be determined as
\begin{equation}
    C_{\textrm{SSC}}(k_i,k_j) = C_{\textrm{small}}(k_i,k_j) + \sigma_b^2 \frac{\partial P(k_i)}{\partial \delta_b} \frac{\partial P(k_j)}{\partial \delta_b}\,, \label{eqn_CSSC}
\end{equation}
where $\sigma_b^2$ is the variance of the background density. If the survey window is large enough that the background density corresponds to a mode in the linear regime, $\sigma_b^2$ can be computed as
\begin{equation}
    \sigma_b^2 = \frac{1}{(2\pi)^3} \int |\tilde{W}^2(kx)|P_{\textrm{lin}}(k)d^3k\,, \label{eqn_sigmab2}
\end{equation}
where $\tilde{W}(kx)$ is the Fourier transform of the survey window function and $P_{\textrm{lin}}(k)$ is the linear power spectrum.

Measuring the power spectrum of a simulation with nonzero $\delta_b$ carries with it the question of whether to normalize the power spectrum relative to this ``local'' average density or relative to the ``global'' average set by the cosmological parameters. To distinguish between quantities using these different normalizations, we adopt the ``local'' and ``global'' subscripts, respectively. The power spectrum measured relative to the global mean density can be related to that of the local mean density by
\begin{equation}
    P_{\textrm{global}}(k) = P_{\textrm{local}}(k)\left( 1 + \delta_b \right)^2\,.
\end{equation}

\subsection{Running separate universe simulations}
To capture the super-sample covariance effect in an ensemble of simulations, we need to be able to run cosmological simulations with non-zero $\delta_b$. This problem has been an active field of study for many years, even before being contextualized by its application for computing super-sample covariance \cite{Hamilton_2006,Desjacques_2018,Sirko_2005,Mcdonald_2003,Goldberg_2004,Martino_2009,Gnedin_2011,Baldauf_2011,Takada_2013,Li_2014,Wagner_2015,Dai_2015,Tormen_1996,Cole_1997,Zhai_2022}. While multiple methods to do so exist, the fundamental principle of each method is the same: the change in average density of the simulation can be interpreted as though the simulation was a separate universe with different cosmological parameters compared to a fiducial background universe with $\delta_b = 0$.

\subsubsection{Sirko method}

Following the derivations used in Ref.~\cite{Sirko_2005}, the change in $\delta_b$ can be interpreted directly as a perturbation in the matter density parameter $\Omega_m$ in the simulation. The matter density parameter of the separate universe can be related to that of the fiducial cosmology by
\begin{equation}
\frac{\Omega_{m,\textrm{uni}}H_{0,\textrm{uni}}^2}{a_{\textrm{uni}}^3(t)}\left(1+\delta_b\right) = \frac{\Omega_{m,\textrm{box}}H_{0,\textrm{box}}^2}{a_{\textrm{box}}^3(t)}\,, \label{eqn_SirkoOmH2}
\end{equation}
where $H_0$ is the Hubble parameter and $a(t)$ is the scale factor. The subscripts ``box'' and ``uni'' refer to parameters in the separate universe and fiducial cosmology, respectively. This in turn modifies other input cosmological parameters for the separate universe
\begin{eqnarray}
    a_{\textrm{box}} = a_{\textrm{uni}}\left( 1 - \frac{\delta_b}{3} \right),\\
    H_{0,\textrm{box}} = H_{0,\textrm{uni}} \frac{1}{1+\phi},\\
    \Omega_{m,\textrm{box}} = \Omega_{m,\textrm{uni}} (1 + \phi)^2,\\
    \Omega_{\Lambda,\textrm{box}} = \Omega_{\Lambda,\textrm{uni}} (1 + \phi)^2\,,
\end{eqnarray}
where
\begin{equation}
    \phi = \frac{5\Omega_{m,\textrm{uni}}}{6}\frac{\delta_b}{D(1)}\,,
\end{equation}
and $D(1)$ is the linear growth factor at scale factor $a=1$.

\subsubsection{Spherical collapse method}
Alternatively, one can interpret a change in $\delta_b$ by considering how a change in the average density within the simulation volume would perturb the curvature of space within. Our description of this approach follows the notation presented in Refs.~\cite{Percival_2005,Zhai_2022}. We can write the Friedmann equation of our $N$-body simulation as:
\begin{equation}
    \frac{1}{H_{0,\textrm{uni}}^2}\left(\frac{da_{\textrm{box}}}{dt}\right)^2 = \frac{\Omega_{m,\textrm{uni}}}{a_{\textrm{box}}} + \epsilon_{\textrm{box}} + \Omega_{\Lambda,\textrm{uni}} a_{\textrm{box}}^2\,,
    \label{eqn_SC_Friedmann}
\end{equation}
where $\epsilon_{\textrm{box}}$ is the perturbed curvature of the box, given by:
\begin{equation}
    \epsilon_{\textrm{box}} = \Omega_{k,\textrm{uni}} - \frac{5\Omega_{m,\textrm{uni}}\delta_b}{3D_0} = \Omega_{k,\textrm{uni}} - 2\phi\,.
    \label{eqn_SC_epsilon_p}
\end{equation}
Note that here $\epsilon_{\textrm{box}} \neq 1 - \Omega_{m,\textrm{uni}} - \Omega_{\Lambda,\textrm{uni}}$. If we re-normalize Equation~\ref{eqn_SC_Friedmann} by dividing both sides by $1 - 2\phi$ and define a new set of renormalized cosmological parameters
\begin{align}
    H_{0,\textrm{box}} &= H_{0,\textrm{uni}}\sqrt{1-2\phi}\nonumber\\
    \Omega_{m,\textrm{box}} &= \frac{\Omega_{m,\textrm{uni}}}{1-2\phi}\nonumber\\
    \Omega_{k,\textrm{box}} &= \frac{\epsilon_{\textrm{box}}}{1-2\phi}\nonumber\\
    \Omega_{\Lambda,\textrm{box}} &= \frac{\Omega_{\Lambda,\textrm{uni}}}{1-2\phi}\,,
    \label{eqn_SC_renormalize}
\end{align}
then the Friedmann equation becomes
\begin{equation}
    \frac{1}{H_{0,\textrm{box}}}\left(\frac{da_{\textrm{box}}}{dt}\right)^2 = \frac{\Omega_{m,\textrm{box}}}{a_{\textrm{box}}} + \Omega_{k,\textrm{box}} + \Omega_{\Lambda,\textrm{box}}a_{\textrm{box}}^2\,.
    \label{eqn_SC_Friedmann_renorm}
\end{equation}
Under this renormalization of the cosmological parameters, $\Omega_{k,\textrm{box}} = 1 - \Omega_{m,\textrm{box}} - \Omega_{\Lambda,\textrm{box}}$, which makes it useful when working with $N$-body codes that infer the value of $\Omega_k$ from the input values of $\Omega_m$ and $\Omega_\Lambda$. This form of the perturbed cosmological parameters is consistent with those found in Refs.~\cite{Desjacques_2018,Baldauf_2011,Wagner_2015}.

\subsection{Addition method}
Once the perturbed cosmological parameters are computed so that we can model the evolution of a patch of the universe with a particular $\delta_b$, there are two possible methods to compute the super-sample covariance effect. The first method, the ``addition'' method, involves computing the power spectrum derivative (the last term in Equation~\ref{eqn_CSSC}). We can do this by generating pairs of simulations, one having $\delta_b > 0$ and the other $\delta_b < 0$. The power spectrum derivative can then be computed from the finite difference of the measured power spectra from the pair of simulations. In principle, this method only requires one pair of simulations to determine the super-sample effect, but typically several realizations are averaged together to decrease the stochasticity in the measured power spectrum derivative. The size of the separate universe simulations used to calculate this response is largely unimportant; the measured power and its response to the background mode does not depend on simulation volume, so as long as the separate universes are large enough to capture the scales of interest they can be used to calculate the super-sample effect within arbitrary survey volumes.

\subsection{Ensemble method}
The second method, the ``ensemble'' method, directly incorporates the effects of super-sample covariance into the ensemble of simulations. To do this, each simulation in the ensemble is a separate universe with $\delta_b$ drawn from a Gaussian distribution with variance $\sigma_b^2$. The covariance matrix calculated from these simulations' power spectra will already include the super-sample effect. This method has the advantage of requiring fewer simulations than the addition method since it requires no additional simulations to be run for the purpose of computing the power spectrum derivative.

\section{Comparison of SSC models} \label{Sec_CompareSSCModels}
The different methods of computing the effects of SSC presented thus far have been verified in previous works to recover the SSC correction with reasonable accuracy \cite{Li_2014,Bayer_2022,Wagner_2015,Howlett_2017}. In this section we confirm this match and, for the first time, compare results from the two proposed ways of perturbing parameters to create ensembles and the additive method.

\subsection{Simulation parameters}
To test these different methods, we generated ensembles of approximate non-linear dark-matter simulations using the $N$-body simulation code L-PICOLA \cite{LPICOLA}. L-PICOLA is a parallelized implementation of the COmoving Lagrangian Acceleration method \cite{Tassev_2013}. When running simulations in parallel across multiple CPUs, L-PICOLA divides the simulation volume into slabs along one axis. Each CPU is assigned one slab for which it evolves its particles using the particle-mesh method. At the end of each timestep, particles that have moved outside of their CPU's slab are moved to the correct slab before the next timestep begins. Due to this parallelization, L-PICOLA can run simulations with large numbers of particles quickly and using few timesteps while still accurately modeling nonlinear clustering. We chose L-PICOLA for its ability to model the non-linear scales of the power spectrum very accurately for relatively low computational cost. The simulations are cubic volumes with periodic boundary conditions.

When using L-PICOLA to run separate universe simulations, it is important to change the value of $\sigma_8$ passed to the simulation code. L-PICOLA must be given the value of $\sigma_8$ at redshift $z_{\textrm{box}} = 0$ which it uses to normalize the input linear power spectrum. Since the separate universe has different cosmological parameters, the value of $\sigma_{8,\textrm{box}}$ will differ from that of the fiducial cosmology. This change can be calculated using the growth factors of the two cosmologies
\begin{equation}
    \sigma_{8,\textrm{box}} = \sigma_{8,\textrm{uni}} \frac{D^2(a_{\textrm{sync}},\Omega_{m,\textrm{uni}},\Omega_{\Lambda,\textrm{uni}})}{D^2(1,\Omega_{m,\textrm{uni}},\Omega_{\Lambda,\textrm{uni}})}\frac{D^2(1,\Omega_{m,\textrm{box}},\Omega_{\Lambda,\textrm{box}})}{D^2(a_{\textrm{sync}},\Omega_{m,\textrm{box}},\Omega_{\Lambda,\textrm{box}})}, \label{eqn_sigma8box}
\end{equation}
where $a_{\textrm{sync}}$ is an early scale factor (typically $a = 0.001$) at which the simulations are synchronized and their power spectra are identical.

Following Ref.~\cite{Li_2014}, care must be taken when handling units in the input and outputs of the simulation codes being used. When running separate universe simulations using L-PICOLA, the code works in units of $h_{\textrm{box}}^{-1}$ Mpc. This requires the initial linear power spectrum to be converted to these units and the output particle positions to be converted back to $h_{\textrm{uni}}^{-1}$ Mpc units before computing the power spectrum.

Since the scale factor $a_{\textrm{box}}$ is different for the separate universe simulations, it is convenient to set their comoving sizes such that, at the output epoch, they all have the same proper sizes. This can be achieved by modifying the size of each simulation to be 
\begin{equation}
    L_{\textrm{box}} = L_{\textrm{uni}}\frac{a_{\textrm{uni}}}{a_{\textrm{box}}}\frac{h_{box}}{h_{uni}}.
\end{equation}
The particle positions can then be converted into proper units before the power spectrum is computed. This ensures that the same physical scales are being compared between the separate universes.

\subsection{Comparison of parameter choices} \label{Sec_Comparison_of_params}

To demonstrate the significance of SSC and compare the effectiveness of the Sirko and SC parameter choices in recovering the SSC effect, we ran ensembles with the following parameters using L-PICOLA:
\begin{itemize}
    \item \textit{Small boxes}: 9728 $L=625\ h^{-1}$ Mpc simulations with $N=256^3$ particles and mesh-grid cells. These simulations have identical cosmological parameters and are used to compute the covariance matrix without any SSC correction.
    \item \textit{Sub-boxes}: 8 $L=5000\ h^{-1}$ Mpc, $N=2048^3$ simulations with identical cosmological parameters. Each simulation was subdivided into 512 sub-boxes with $L=625\ h^{-1}$ Mpc, resulting in a total of $4096$ sub-boxes. These simulations innately include the effect of SSC and will serve as the benchmark that the other ensembles will try to match the covariance.
    \item \textit{Sirko addition method}: $128\ L=625\ h^{-1}$ Mpc, $N=256^3$ simulations, where the first set of 64 simulations has been generated with cosmological parameters corresponding to $\delta_b=-0.01$, and the remaining 64 generated with $\delta_b=0.01$. The two sets of simulations are run with the same set of 64 initial seeds. These simulations are used to compute the power spectrum derivative needed for the SSC term in Equations~\ref{eqn_CSSC}. The no-SSC term is computed using the “Small boxes” ensemble covariance.
    \item \textit{SC addition method}: $128\ L=625\ h^{-1}$ Mpc, $N=256^3$ simulations identical in setup to the ``Sirko addition method'' simulations except using the SC method parameters from Equation~\ref{eqn_SC_renormalize} 
\end{itemize}

Each ensemble has fiducial background cosmological parameters given by Table~\ref{table_cosmo_params} and each mock run to an output redshift corresponding to $a_{\textrm{uni}}=1$. They were evolved from an initial redshift of $z_i = 100$ using 60 logarithmically spaced time steps. The power spectra were computed from the output particle catalogues using Nbodykit \cite{Hand2018_nbodykit} without any shot noise subtraction. The particles were painted to an overdensity field using a PCS resampler to accurately measure the power spectrum to the Nyquist frequency. Only the dark matter power spectrum is considered in this work as the SSC effect on smaller-scale modes is largely washed out by shotnoise in galaxy power spectra. The power spectra and covariance matrices were calculated using bins of width $\Delta k = 10\pi / 625\ \sim 0.05$ $h_{uni}$ Mpc$^{-1}$, spanning a range of values from $0 < k < 1.3\ h_{uni}$ Mpc$^{-1}$. The uncertainty on the recovered covariance was estimated based on the matrices being drawn from the Wishart distribution.

\begin{table}
\centering
    \begin{tabular}{|c|c|}
    \hline
    Cosmological parameter & Value \\
    \hline
    h & 0.6736 \\
    $\Omega_{b}h^2$ & 0.02237 \\
    $\Omega_{cdm}h^2$ & 0.12 \\
    $n_s$ & 0.9649 \\
    $\sigma_8$ & 0.8111 \\
    \hline
    \end{tabular}
    \caption{Flat $\Lambda$CDM parameters chosen for the background universe.}
    \label{table_cosmo_params}
\end{table}

\begin{figure*}
    \includegraphics[width=\linewidth]{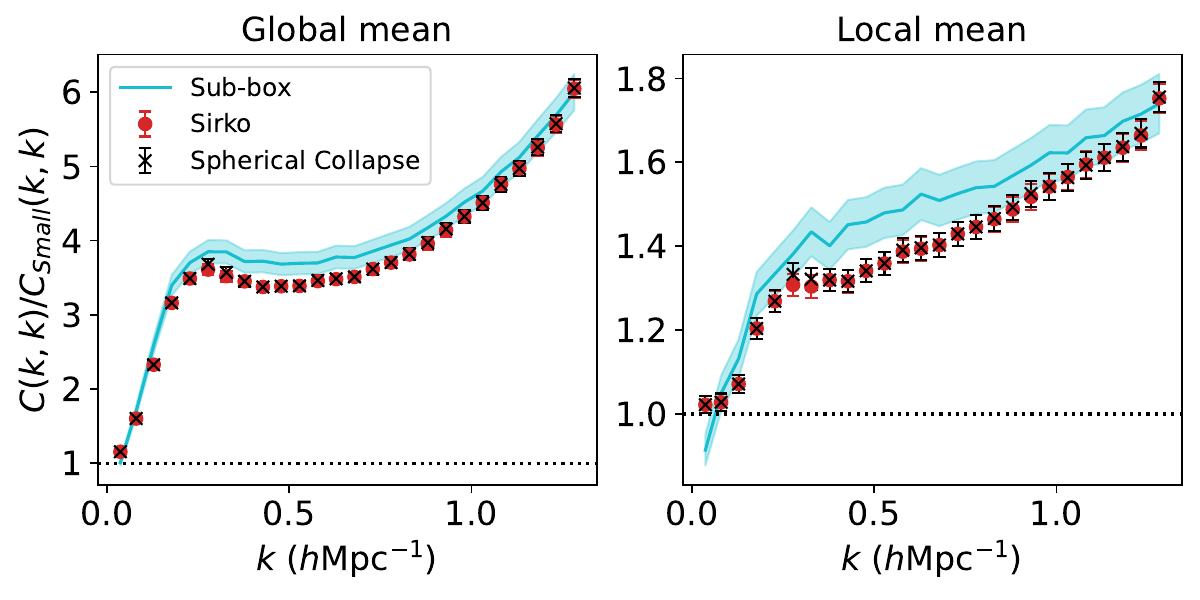}
    \caption{Comparison of Sirko (red $\times$) and Spherical Collapse (SC) (green plus) method variance, divided by the variance from the small boxes (without super-sample covariance). The blue curve and shaded region show the variance from the sub-boxes of the 5000 $h^{-1}$ Mpc box and its 2$\sigma$ confidence interval, which we take as the true covariance including SSC.
    The left and right panels show the variance relative to the global (i.e.\ relative to the 5000 $h^{-1}$ Mpc box) and local mean density (i.e.\ relative to each 625 $h^{-1}$ Mpc sub-box), respectively. The error bars show the 95\% confidence interval.}
    \label{fig_SirkoSC}
\end{figure*}
\begin{figure*}
    \includegraphics[width=\linewidth]{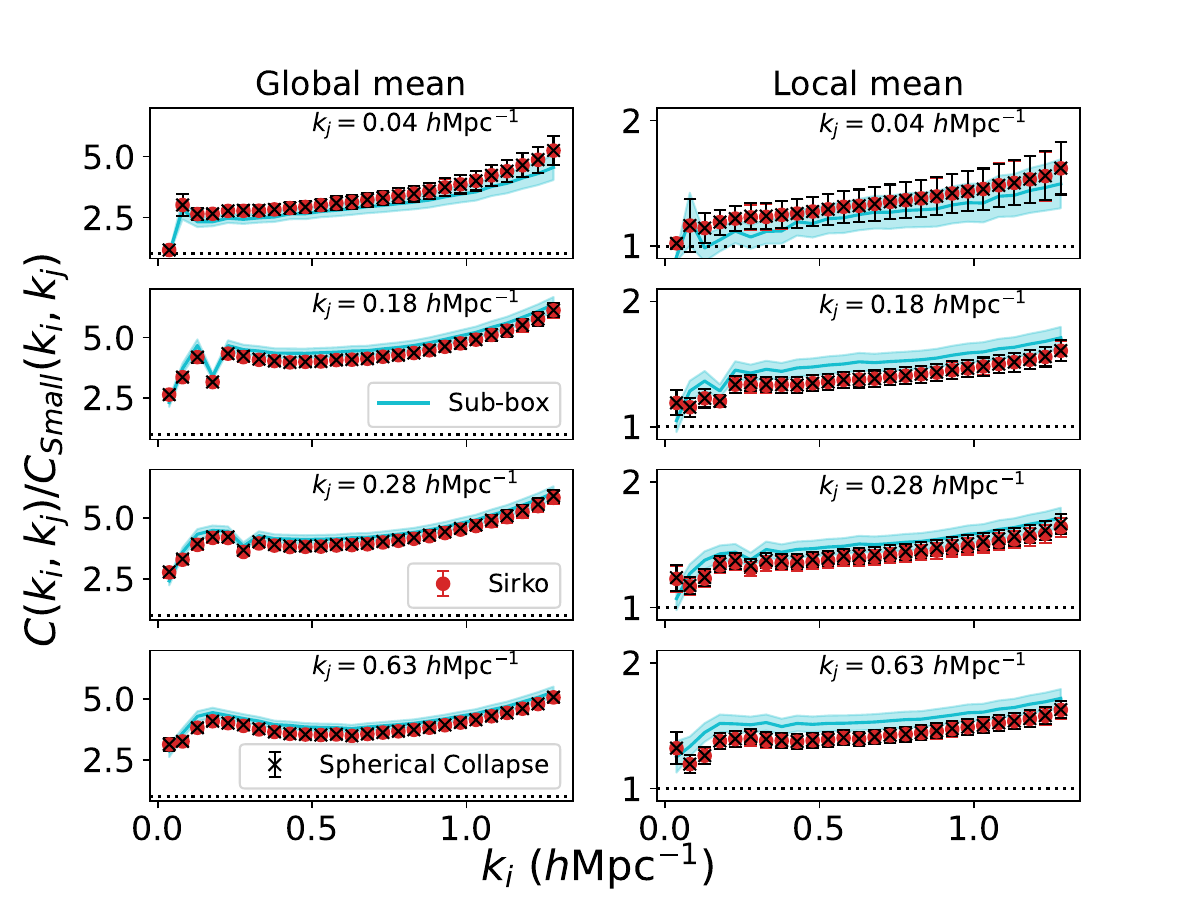}
    \caption{Comparison of select off-diagonal elements of the covariance matrix, showing the first, fifth, tenth, and fifteenth bins for $k_j$. Black $\times$ represent SC addition method mocks, and red circles represent Sirko addition method mocks. The blue curve and shaded region show the covariance from the sub-boxes of the 5000 $h^{-1}$ Mpc box and its 2$\sigma$ confidence interval. The left and right panel show the global and local mean results, respectively.}
    \label{fig_SirkoSCOD}
\end{figure*}

Figure~\ref{fig_SirkoSC} shows the comparison between the variance calculated using the Sirko and SC method. The SSC effect on the variance relative to the global mean is to increase it by over 100$\%$ for modes of $k\sim 0.1 \ h$ Mpc$^{-1}$ or greater compared to simulations without SSC. The effect grows larger as $k$ increases, reaching a maximum of almost 500$\%$ at $k = 1.28 \ h$ Mpc$^{-1}$. This increase in variance is primarily driven by the normalization of the power spectrum with different $\delta_b$. In the local mean case, which is more observationally relevant for galaxy surveys, the SSC contribution is still significant, reaching almost 80$\%$ at $k=1.28 \ h$ Mpc$^{-1}$. In Figure~\ref{fig_SirkoSCOD}, the effect of SSC on the off-diagonal terms of the covariance matrix can be seen to be substantial even at low $k_j$.

Both the Sirko and the SC methods perform well in modeling the SSC effect and were able to recover the same covariance as the sub-boxes to within 10$\%$ or better for most bins in both the local and global cases. The two methods were consistent with one another, with the SC method only marginally outperforming the Sirko method in some of the bins. The variance from both methods was biased low compared to the sub-boxes, and most bins fell outside of the 95 percent confidence interval of the sub-box variance. This bias could be a product of using the addition method, which itself is only a first order approximation of the ensemble method.

\subsection{Comparing addition and ensemble methods}

To compare the performance of the addition and ensemble methods, we ran a new ensemble of simulations we refer to as ``Ensemble SC'', containing 9728 simulations with $L = 625\ h^{-1}$ Mpc containing $(256)^3$ particles and with background overdensity drawn from a Gaussian distribution as prescribed by the ensemble method. The cosmological parameters for each simulation were computed using the SC method and used the same fiducial background cosmology given in Table~\ref{table_cosmo_params}.

\begin{figure*}
    \includegraphics[width=\linewidth]{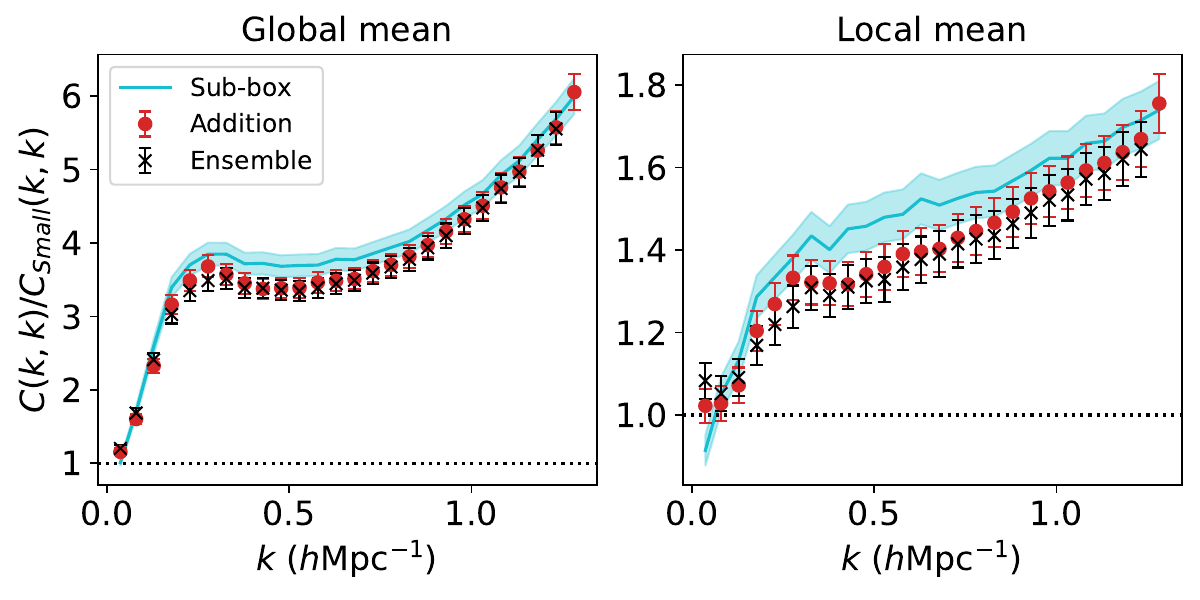}
    \caption{Comparison of SC addition method (red circles) and ensemble method (black $\times$) variance over small mock variance. The blue curve and shaded region show the variance from the sub-boxes and its 2$\sigma$ confidence interval. The left and right panels show variance relative to the global and local mean density, respectively.}
    \label{fig_SCaddens}
\end{figure*}

\begin{figure*}
    \includegraphics[width=\linewidth]{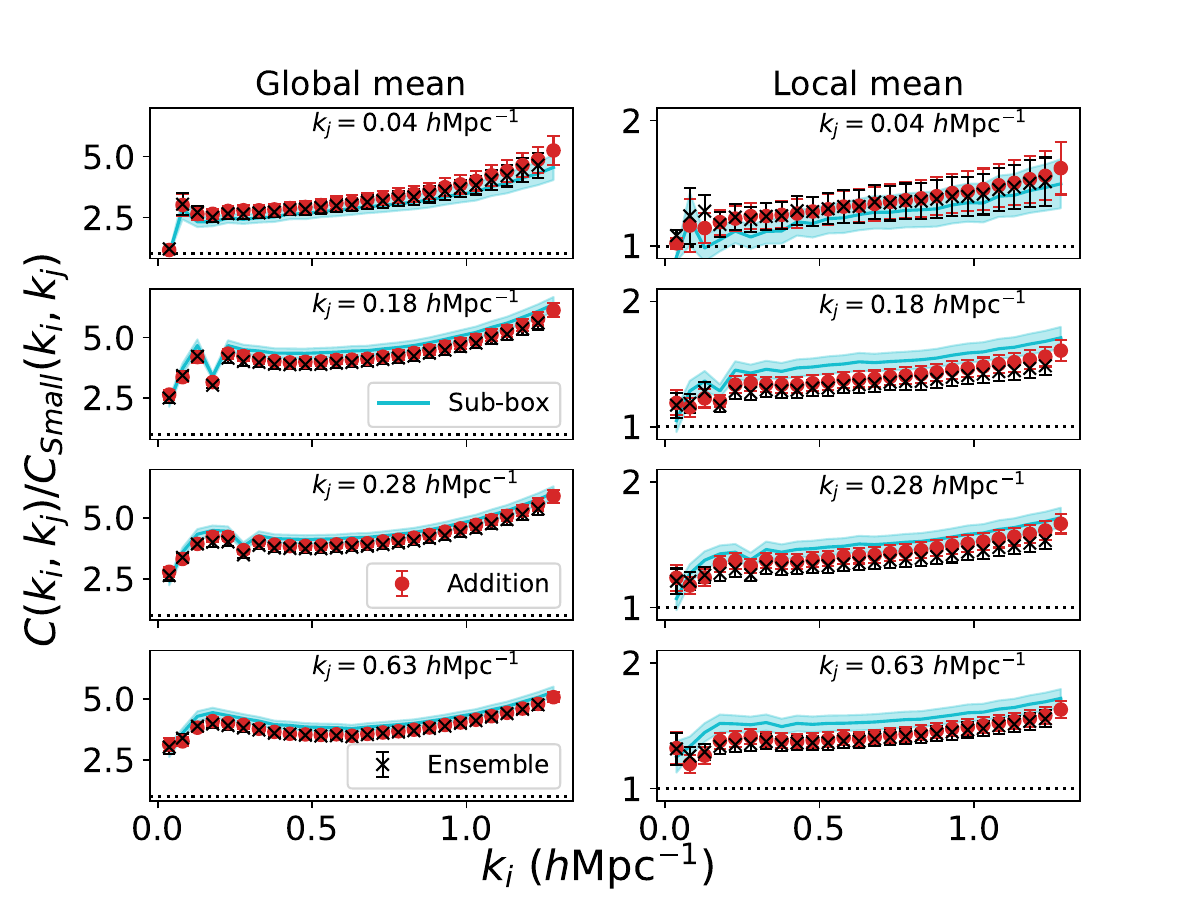}
    \caption{Comparison of select off-diagonal elements of the covariance matrix, showing the first, fifth, tenth, and fifteenth bins for $k_j$. Red circles represent SC addition method mocks, and black $\times$ represent SC ensemble method mocks. The blue curve and shaded region show the variance from the sub-boxes and its 2$\sigma$ confidence interval. The left and right panels show variance relative to the global and local mean density, respectively.}
    \label{fig_SCaddensOD}
\end{figure*}

For both the global and local normalizations of the power spectra, Figures~\ref{fig_SCaddens} and~\ref{fig_SCaddensOD} show that the addition and ensemble methods are consistent with one another for both on-diagonal and off-diagonal terms of the covariance matrix. The ensemble method estimates the covariance a few percent higher than the addition method in most $k$ bins, causing it to match more closely to the sub-box covariance. This is because the addition method is a first order approximation of the ensemble method. While this first order approximation is good enough to recover the majority of the SSC effect, the higher order terms make a few percent improvement to the overall accuracy.

\subsection{Comparison with previous studies}

\begin{figure*}
    \includegraphics[width=\linewidth]{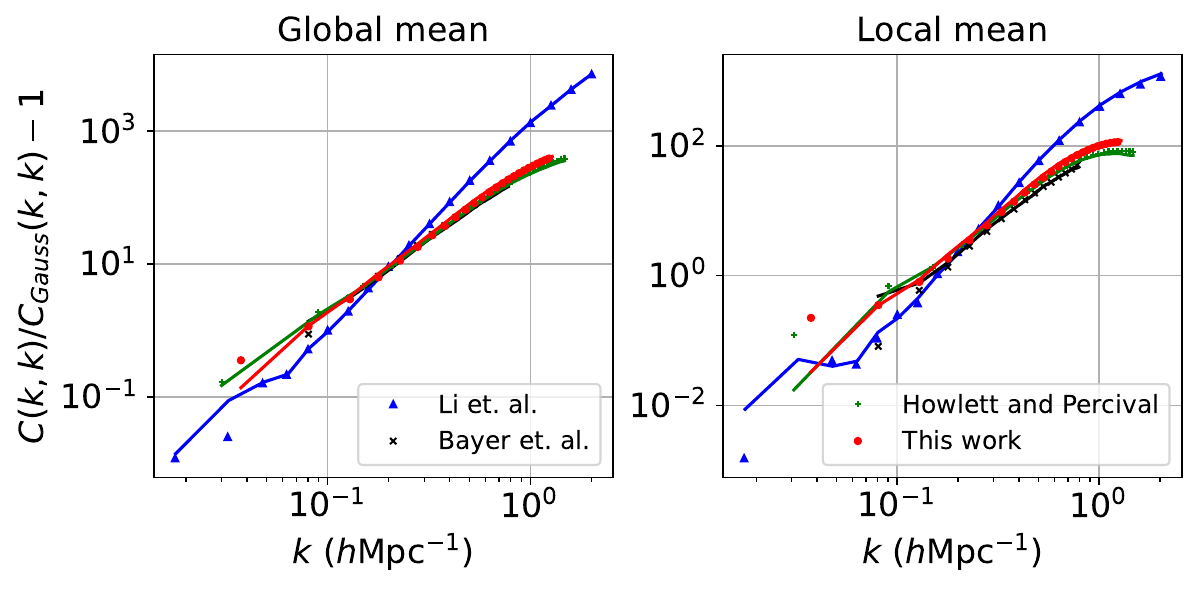}
    \caption{Variance over the Gaussian expectation of sub-box (solid lines) and SU (markers) simulations from Li et~al. Figures 6 and 7 \ \cite{Li_2014} (blue), Bayer et~al. Figure 2 \ \cite{Bayer_2022} (black), Howlett and Percival \cite{Howlett_2017} Figure 3 (green) and this work (red). The left and right panels show global and local mean results respectively. In the left panel, the data of Bayer et~al. is largely obscured by the data of Howlett and Percival.}
    \label{fig_Lifig6}
\end{figure*}

This level of agreement between the sub-box covariance and the different SSC methods is comparable to that recovered in previous studies \cite{dePutter_2012,Hu_2001,Li_2014,Takahashi_2009,Bayer_2022,Hamilton_2006,Howlett_2017}. To compare our results to these studies, we computed the variance over the Gaussian expectation $C_{\textrm{Gauss}}$ for our power spectra and the data from Refs.~\cite{Bayer_2022} and \cite{Howlett_2017} to match the format of figures 6 and 7 from Ref.~\cite{Li_2014}. The results of this are shown in Figure~\ref{fig_Lifig6}. The SC ensemble method we used was able to capture the effect of SSC with a comparable level of success to the results from Refs.~\cite{Li_2014}, \cite{Bayer_2022} and \cite{Howlett_2017}, each of which also found agreement between sub-box and SU variance to within 10 percent or better. The agreement between this work, Ref.~\cite{Bayer_2022}, and Ref.~\cite{Howlett_2017} is especially good. Differences in the measured covariances between studies can be attributed to the different choices of fiducial cosmologies and simulation codes used. In particular, the order of magnitude difference in covariance measured at high $k$ between the results of Ref.~\cite{Li_2014} and the other works is likely driven by the fact that the former used the simulation code L-Gadget2 \cite{Springel_2005}. The full $N$-body (Tree)-PM method used by L-Gadget2 is able to more accurately model the power spectrum at high $k$ than fast simulation codes such as L-PICOLA and FastPM at the cost of requiring more computation time. This suggests that fast $N$-body codes under-estimate the full non-Gaussian covariance at $k > 0.2\ h$ Mpc$^{-1}$, where full $N$-body codes are needed for accurate covariance estimation. Regardless of simulation code, the internally consistent ability to recover small-scale covariance demonstrates the effectiveness of modeling super-sample covariance.

\section{Volume scaling of the covariance matrix} \label{Sec_VolumeScaling}

We now consider how well we can use volume scaling to speed up the covariance matrix calculation. 

\subsection{Motivation: reducing the error on the covariance matrix} \label{sec_reducingErrorOnCovariance}
When estimating a covariance from a number of simulations $N$ much greater than the number of bins in the power spectrum, the error on the covariance matrix scales as
\begin{equation}
    \Delta C \propto \sqrt{\frac{1}{N}}\,. \label{eqn_deltaC1}
\end{equation}
This scaling has been shown to hold even for non-linear simulations \cite{Takahashi_2009,Takahashi_2011}. The error on the covariance matrix also scales proportionally to the covariance itself, which in turn scales inversely with the survey or simulation volume $V$ in the absence of super-sample covariance. Equation~\ref{eqn_deltaC1} can be modified to reflect this
\begin{equation}
    \Delta C \propto \sqrt{\frac{1}{N}}\frac{1}{V}\,.
\label{eqn:volume_scaling}
\end{equation}

Using the fact that both terms in Equation~\ref{eqn_CnoSSC} scale inversely with volume, we can estimate the covariance within a large survey window with volume $V_L$ by running an ensemble of small simulations with volume $V_S$, computing the covariance matrix, and re-scaling it by a factor of $V_S/V_L$. In a given amount of computation time, this re-scaled covariance from the small volume mocks $C_{S,\textrm{scaled}}$ will have reduced error compared to that from the full volume mocks $C_L$
\begin{equation}
    \Delta C_{S,\textrm{scaled}} \propto \sqrt{\frac{N_L}{N_S}}\Delta C_L,
\end{equation}
where $N_L$ and $N_S$ are the number of large and small volume mocks able to be run in the given computation time.

While this volume scaling technique sees greater improvement on the precision of the covariance matrix estimation as simulation volume decreases, the volume scaling in Eq.~\ref{eqn:volume_scaling} must be corrected for super-sample covariance to maintain accuracy. When doing this, the simulations cannot be made arbitrarily small: the simulations must be large enough to contain modes at scales of interest for measurements of the power spectrum to be made. Additionally, the small scale modes in these small volume simulations evolve without the coupling to large scale modes that would be present in full volume simulations, causing the covariance on small scales to be biased compared to the ``true'' covariance. The super-sample covariance term also scales as a complex function of the simulation volume, so care must be taken when adding it back into the simulations. For the addition method, $\sigma_b$ must be calculated to correspond to the full survey volume and the SSC term added to the covariance after volume scaling has taken place. The volume of the simulations used to compute the power spectrum derivative does not matter since the power spectrum derivative is expected to be volume independent. The total covariance for the survey volume $C_{L}$ is then computed as
\begin{equation}
    C_{L}(k_i,k_j) = \frac{V_{S}}{V_{L}}C_{S}(k_i,k_j) + \sigma_{b,L}^2\frac{dP(k_i)}{d\delta_b}\frac{dP(k_j)}{d\delta_b}, \label{eqn_volscale_addition}
\end{equation}
where $V_L$ and $V_S$ are the volumes of the survey and the simulations, respectively, and $\sigma_{b,L}^2$ is the variance in background density calculated as in Equation~\ref{eqn_sigmab2} using the large box volume.

For the ensemble method, the variance of the background overdensity of the small simulations must be set to $V_L/V_S \sigma_{b,L}^2$ rather than $\sigma_{b,S}^2$, the actual variance in background overdensity calculated for the small volume. This ensures that, after volume scaling has been applied, the super-sample covariance contribution will be equal to that expected in the large survey volume. Incorrectly using $\sigma_{b,S}^2$ calculated for the small simulation volume will result in a substantial overestimation of the SSC term since it scales with volume more strongly than $1/V$. The techniques of reintroducing super-sample covariance discussed in Section~\ref{Sec_CovEstimation} are only expected to work so long as the super-sample modes fall within the linear regime; if the simulation volume is small enough that non-linear modes are being left out, a higher order correction would be necessary to correctly capture the super-sample covariance.

There is also a correction that needs to be applied arising from the discrete $k$ modes in each bin of the power spectrum. The power spectrum measured in a given $k$ bin is computed from modes
that fit within that bin.
Since only certain discrete
values of $k$ are allowed from the volume of the box and the mesh size used, the mean $k$ mode is not necessarily the centre of the bin. As the volume changes, the modes contained in the bin will also change. This means that the power in this same $k$ bin is measured at a slightly different average $k$ value for the large and small mocks. This also changes the measured covariance in this $k$ bin, with the change becoming more significant at lower $k$. To recover the correct covariance of the large-volume mocks using the volume scaling technique, we apply a corrective factor to the measured power of each of the small-volume mocks $P_S(k)$
\begin{equation}
    P_{\textrm{corr},S}(k) = \frac{\overline{P}_L(k)}{\overline{P}_S(k)} \sqrt{\frac{V_{k,S}}{V_{k,L}}}P_S(k), \label{eqn_binCentreCorrection}
\end{equation}
where $P_{\textrm{corr},S}(k)$ is the corrected small volume power spectrum, $\overline{P}_L(k)$ and $\overline{P}_S(k)$ are the ensemble average power of the large and small volume mocks respectively, and $V_{k,S}$ and $V_{k,L}$ are the $k$-space volumes of the bin of interest in the small and large volume mocks respectively, computed by directly summing the volumes contributed by each mode in that bin. This corrective factor was chosen based on the Gaussian behaviour of the covariance matrix at low $k$ where the bin centering issue is most significant.

\subsection{Super-sample covariance and volume scaling}
To evaluate how well the volume scaling technique is able to recover the covariance matrix of a large volume survey, we generated five ensembles of 9728 simulations each with volumes of $(2500\ h^{-1}$ Mpc$)^3$, $(1250\ h^{-1}$ Mpc$)^3$, $(625\ h^{-1}$ Mpc$)^3$, $(312.5\ h^{-1}$ Mpc$)^3$, and $(156.25\ h^{-1}$ Mpc$)^3$ respectively. The simulations each had zero background overdensity, contain a number density of particles $N/V = (256/625)^3\ h^3$Mpc$^{-3}$ and used cosmological parameters given in Table~\ref{table_cosmo_params}. The covariance matrix of each ensemble was computed and volume scaled to match a survey with volume $(2500\ h^{-1}$ Mpc$)^3$. The super-sample covariance term was computed using the SC parameters with the addition method. The power spectrum derivative was calculated using the \textit{SC addition method} ensemble described in Section~\ref{Sec_Comparison_of_params}. In principle, the power spectrum derivative does not depend on the simulation volume, so it can be calculated using any size simulation. In practice, one must still choose a simulation volume containing enough modes at scales of interest to be measured. In this section, we used simulations with a volume of $(2500\ h^{-1}$ Mpc$)^3$ to calculate the power spectrum derivative. For smaller volume simulations, the bin centreing issue discussed in Section~\ref{sec_reducingErrorOnCovariance} is also present. This effect only significantly changes the power spectrum derivative in low $k$ bins where the SSC contribution is weakest, and we have verified that, for the ensembles considered in this work, this effect is negligible when calculating the derivatives.

\begin{figure*}
    \centering
    \includegraphics[width=\linewidth]{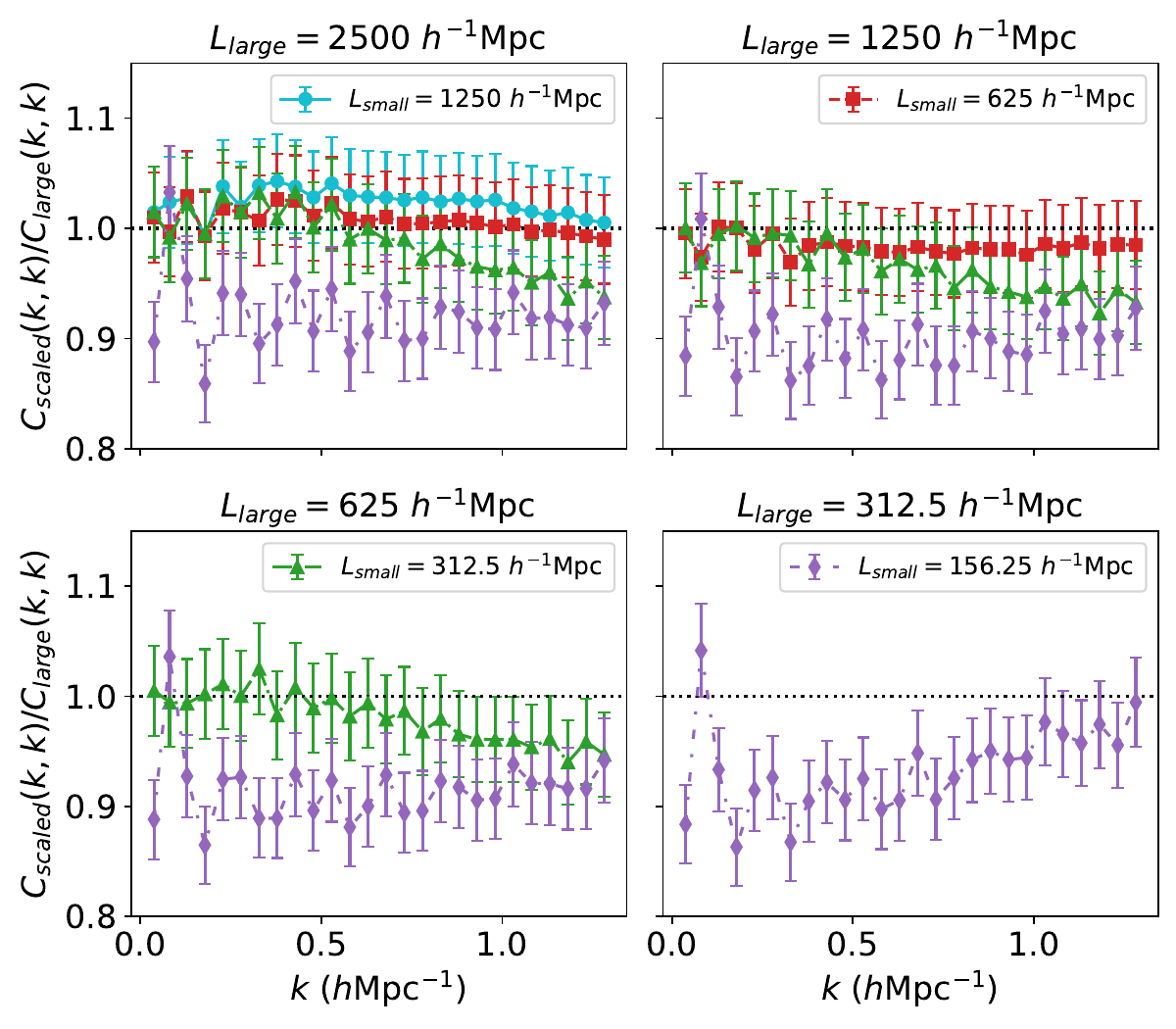}
    \caption{Ratio of volume-scaled small box covariance and large box covariance on-diagonal elements. All covariances are computed relative to the local mean density. The bin centering correction was applied to the power spectra of each ensemble. Each panel corresponds to a different large box volume used as the denominator. Top left: $V_{\textrm{large}} = (2500\ h^{-1}$ Mpc$)^3$. Top right: $V_{\textrm{large}} = (1250\ h^{-1}$ Mpc$)^3$. Bottom left: $V_{\textrm{large}} = 625\ h^{-1}$ Mpc$)^3$. Bottom right: $V_{\textrm{large}} = (312.5\ h^{-1}$ Mpc$)^3$.}
    \label{fig_volscaleOndiagLocal}
\end{figure*}

\begin{figure*}
    \centering
    \includegraphics[width=\linewidth]{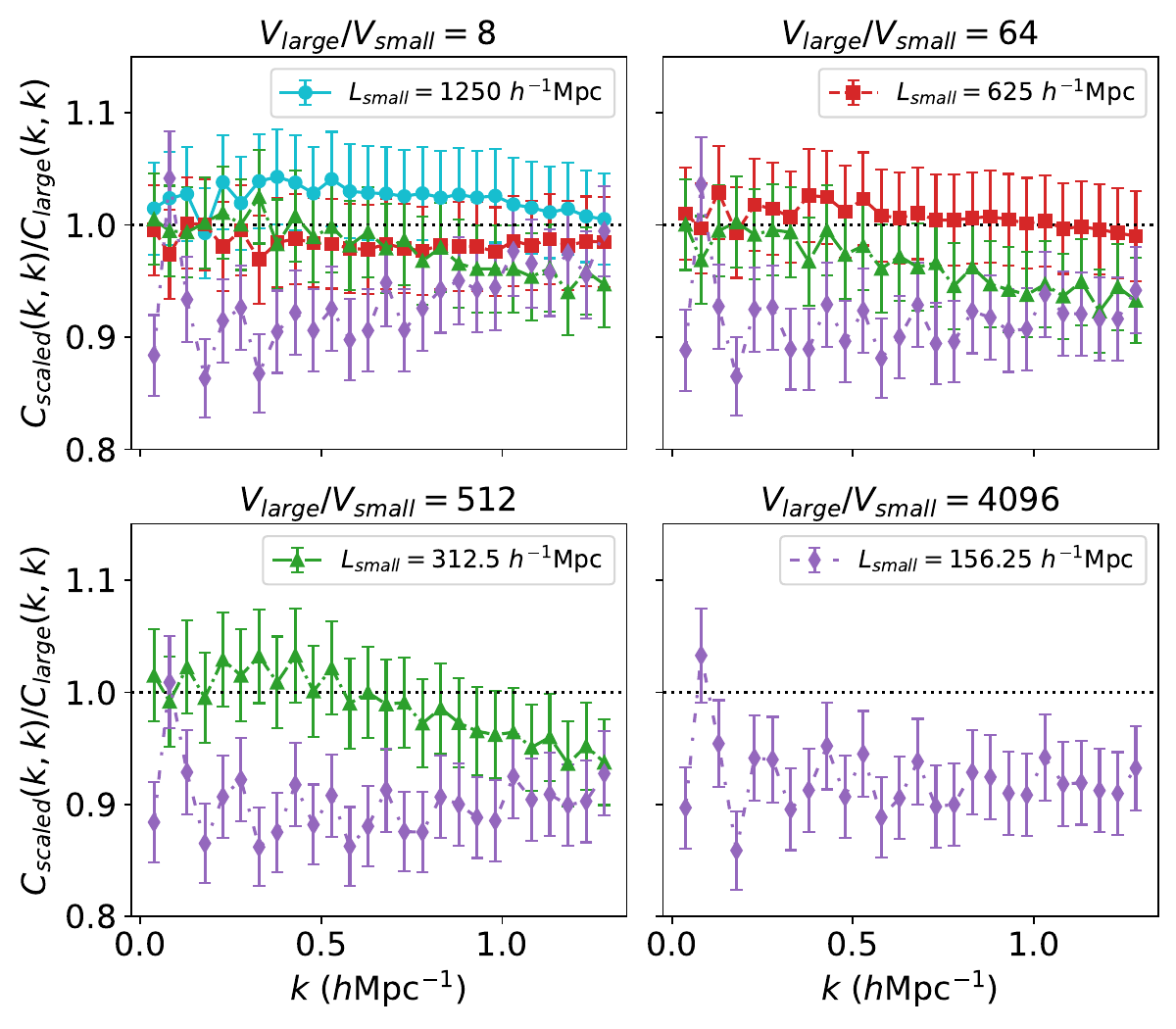}
    \caption{Same as Figure~\ref{fig_volscaleOndiagLocal}, but each panel corresponds to a fixed volume ratio. Top left: $V_S/V_L = 1/8$. Top right: $V_S/V_L = 1/64$. Bottom left: $V_S/V_L = 1/512$. Bottom right: $V_S/V_L = 1/4096$.}
    \label{fig_volscaleOndiagLocalByRatio}
\end{figure*}

\begin{figure}
    \centering
    \includegraphics[width=\linewidth]{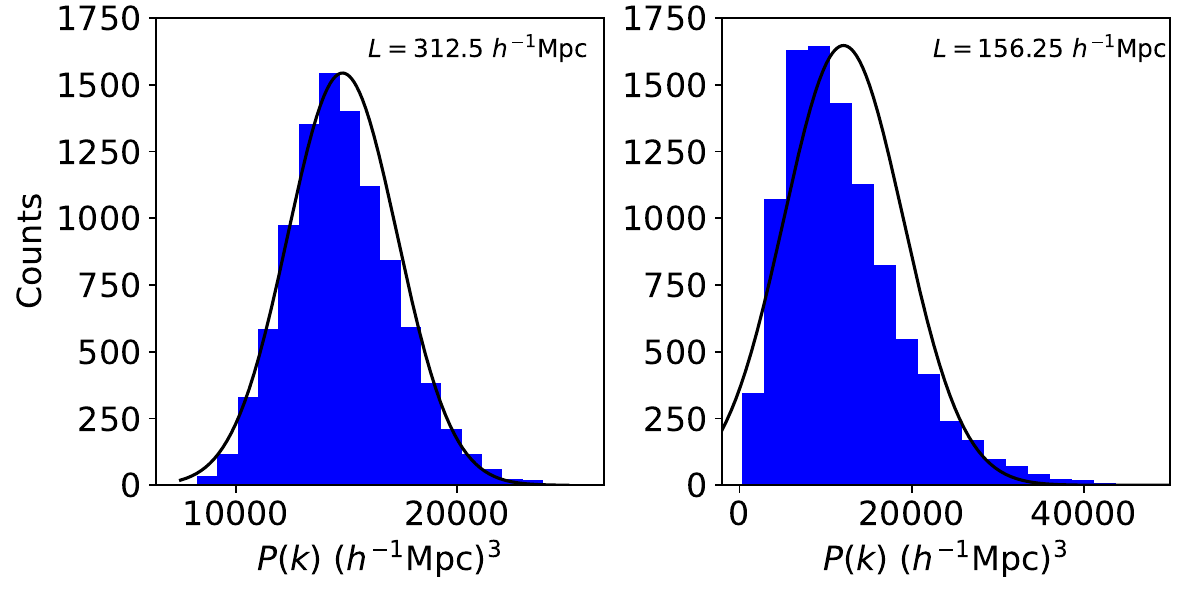}
    \caption{Histogram of power spectrum amplitudes in the $k=0.04\ h$ Mpc$^{-1}$ bin of the $L=312.5\ h^{-1}$ Mpc ensemble (left panel) and $L=156.25\ h^{-1}$ Mpc ensemble (right panel). The solid black curve shows a Gaussian distribution with mean and variance matching those computed from the power spectra of the ensemble. There is a noticeable skewness in the distribution compared to a Gaussian.}
    \label{fig_LowkSkewHistogram}
\end{figure}

\begin{figure*}
    \centering
    \includegraphics[width=\linewidth]{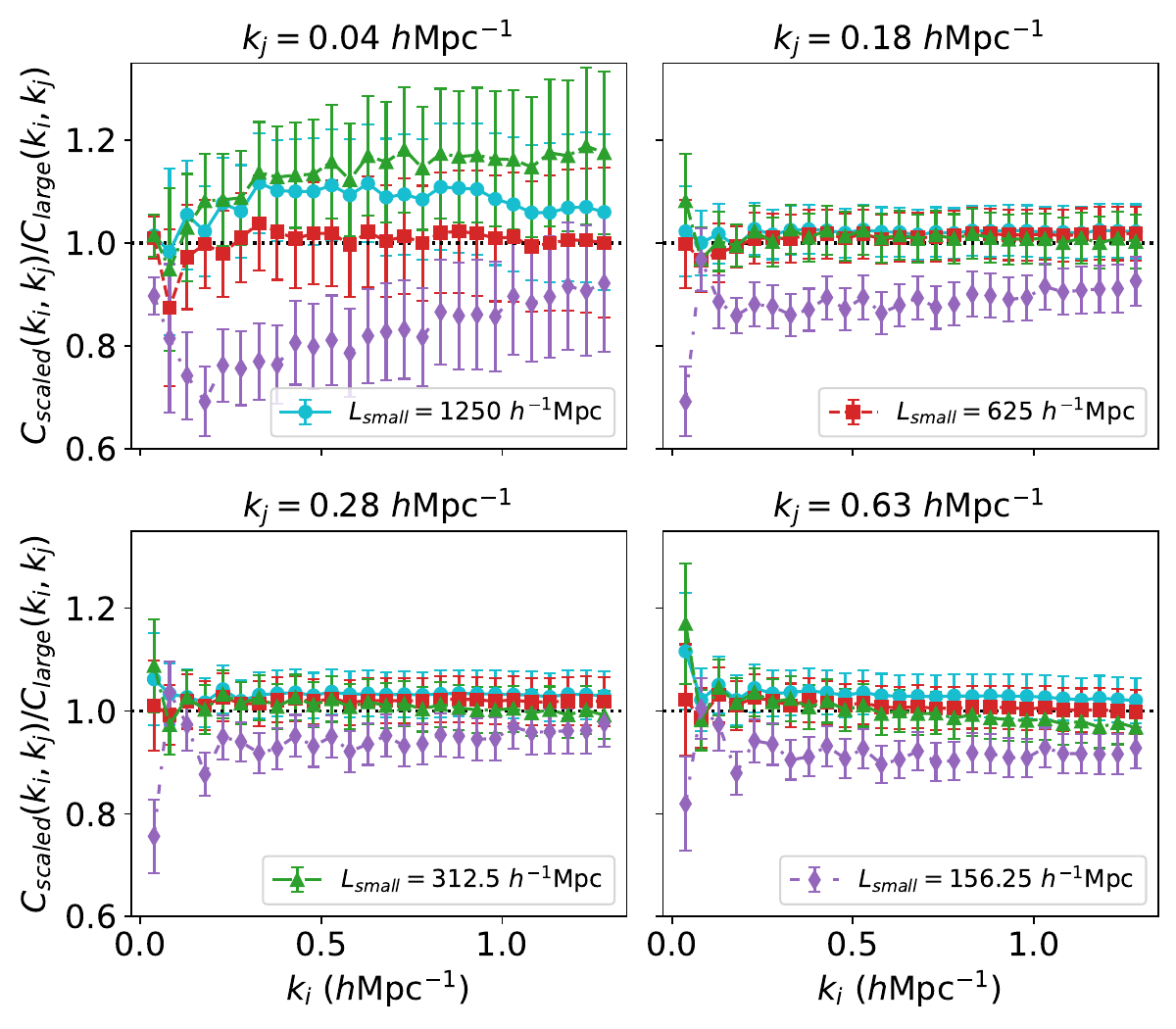}
    \caption{Ratio of volume-scaled small box covariance and $V_{\textrm{large}} = 2500\ (h^{-1}$ Mpc$)^3$ box covariance for select off-diagonal elements. All covariances are computed relative to the local mean density. The bin centering correction was applied to the power spectra. Each panel corresponds to a different $k$ bin.}
    \label{fig_volscaleOffdiag2500local}
\end{figure*}

Figures~\ref{fig_volscaleOndiagLocal},~\ref{fig_volscaleOndiagLocalByRatio}, and~\ref{fig_volscaleOffdiag2500local} show the results of using volume scaling to recover the large volume covariance matrix. We plot the ratio of the volume scaled small mock covariance over the ``large'' mock covariance. We find that, for all volumes tested with the exception of $V = (156.25\ h^{-1}$ Mpc$)^3$, the scaled covariance matches the true covariance to within 3$\%$ or better on almost all scales. This level of agreement is comparable to that of semi-analytic covariance models designed for use with DESI Y1 data \cite{Rashkovetskyi_2024}. The volume-scaled covariance ratio depends on both the volume ratio and the large mock volume, with the covariance ratio decreasing as the volume ratio decreases and as the large mock volume decreases. In Figure~\ref{fig_volscaleOndiagLocal}, the large mock covariance is generally more accurately recovered in the case of volume ratios closer to one. This can be explained by the fact that the covariance calculated from the small volume simulations is missing a contribution from modes with sizes between that of the large and small box. The SSC term added accounts for the coupling of modes larger than the large box, but does not capture the coupling of these missing intermediate-scale modes. The greater the volume ratio between the mocks, the more of these intermediate modes are missing, resulting in a greater underestimation of the covariance. While there is no simple method to re-introduce the effects of these missing modes into the covariance, using tidal fields to model the contribution to SSC by anisotropic background density perturbations could potentially capture some of this missing covariance \cite{Dai_2015}.

Figure~\ref{fig_volscaleOndiagLocalByRatio} shows that the large mock covariance is better recovered as the volume of the large mocks increases independently of the volume ratio. The accuracy of the volume-scaled covariance depends more strongly on the large mock volume than the volume ratio of the mocks. This dependence on large mock volume can similarly be explained by missing intermediate modes in the small mocks. The strength of coupling between modes in a survey or simulation depends on the size of the modes in question (see e.g.~Figure~\ref{fig_SirkoSC}). Figure~\ref{fig_SirkoSCOD} shows that the coupling between modes increases as $k$ increases. This means that the intermediate modes missing from volume scaling from $L=1250\ h^{-1}$ Mpc to $L=2500\ h^{-1}$ Mpc are less correlated with the smaller modes of the power spectrum than those missing when volume scaling from $L=625\ h^{-1}$ Mpc to $L=1250\ h^{-1}$ Mpc, resulting in the former case recovering the covariance more accurately than the latter. Unexpectedly, the $1250/2500$ ratio and the $625/2500$ ratio generally overestimate the covariance matrix after volume scaling despite the missing intermediate mode contribution, though $C_{large}$ still lies within 2$\sigma$ of $C_{scaled}$ in both cases.

The $V = (156.25\ h^{-1}$ Mpc$)^3$ mocks also fail to recover the correct covariance, especially at low $k$. This is due to the low $k$ bins of the smaller volume mocks containing a very small number of bins, causing the power spectra measured in these bins to be significantly non-Gaussian. Figure~\ref{fig_LowkSkewHistogram} demonstrates this in the data from the $k=0.04\ h$ Mpc$^{-1}$ bin of our $L=312.5\ h^{-1}$ Mpc and $L=156.25\ h^{-1}$ Mpc ensemble, where the number of modes in this bin is $N_{\textrm{modes}} = 81$ and $N_{\textrm{modes}} = 6$, respectively. In the linear regime, the power spectrum measured from one realization of the density field is drawn from a Rayleigh distribution. In bins containing a large number of modes, the central limit theorem causes the distribution to be approximately Gaussian. When this approximation breaks down, the covariance matrix can no longer completely describe statistical uncertainty on the power spectrum. Choosing a non-Gaussian shape for the posterior distribution of the cosmological parameters has been shown to produce accurate results when working with such non-Gaussian data \cite{Bond_2000,Smith_2006,Percival_2006,Hamimeche_2008,Kalus_2016,Kalus_2023}.

\begin{figure*}
    \includegraphics[width=\linewidth]{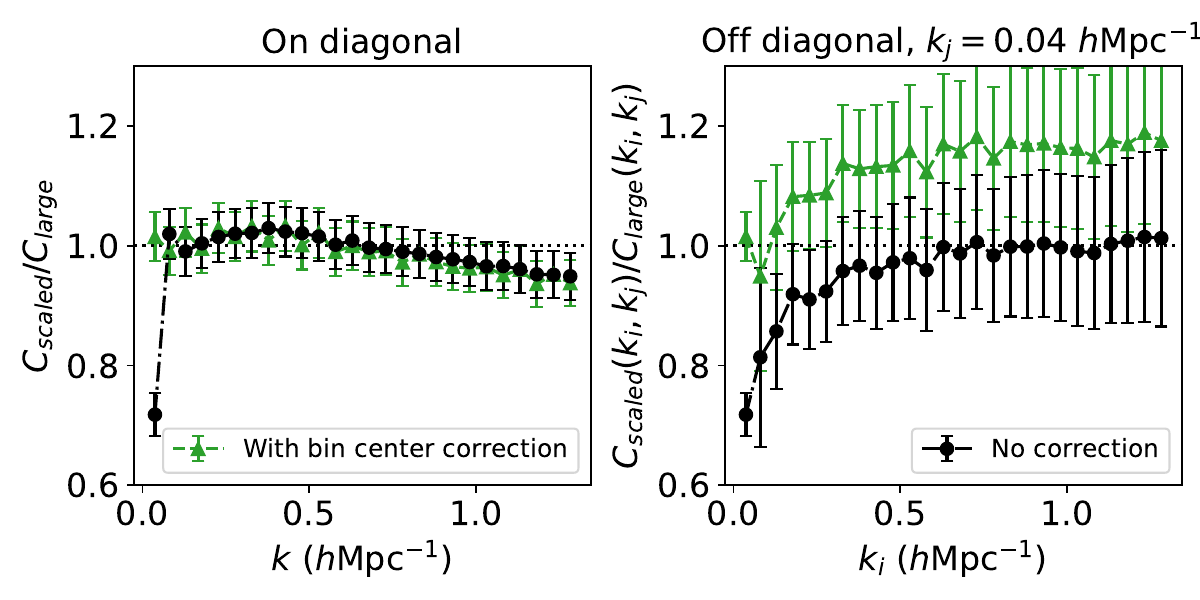}
    \caption{Ratio of volume-scaled $V = (312.5\ h^{-1}$ Mpc$)^3$ covariance and $V = (2500\ h^{-1}$ Mpc$)^3$ covariance shown with the bin centre correction applied (green triangles) and without the correction (black circles). The left panel shows the on-diagonal elements of the covariance matrix, and the left panel shows the off-diagonal elements for $k_i = 0.04\ h$Mpc$^{-1}$ where the correction has the most significant effect.}
    \label{fig_binCorrection}
\end{figure*}

Figure \ref{fig_volscaleOffdiag2500local} shows that the off-diagonal elements of the covariance are also accurately recovered using the volume scaling technique, with most $k$ bins having comparable accuracy to the on-diagonal elements. However, at low $k$ (top left panel in Figure~\ref{fig_volscaleOffdiag2500local}), the volume-scaled covariance deviates significantly from the large mock covariance. This is due to the $k$ bin centre correction assuming the covariance in these bins is Gaussian; the Gaussian expectation for off-diagonal terms of the covariance in the absence of a window function is zero, so the correction in Equation~\ref{eqn_binCentreCorrection} is not expected to model it accurately on these large, linear scales. Figure \ref{fig_binCorrection} demonstrates the effect of this correction on the lowest $k$ bin of one of our ensembles. The diagonal term of the volume-scaled covariance at $k=0.04\ h$Mpc$^{-1}$ is greatly underestimated due to the bin centering issue, but it is accurately recovered once the correction is applied. For the off-diagonal terms of $C(k_i,k_j)$, the uncorrected volume-scaled covariance is still underestimated at low $k$, but more accurately matches the large mock covariance beyond $k \sim 0.2\ h$Mpc$^{-1}$. After correction, the off-diagonal terms become overestimated. This is due to the bin centre correction's inaccurate assumption that the off-diagonal terms of $C(k_i,k_j)$ scale with $P(k)$ and $V_{k}$ similarly to the Gaussian piece of the covariance, which dominates on-diagonal. In actuality, the behaviour of the off-diagonal terms is governed by the trispectrum (Equation~\ref{eqn_trispectrum}) and SSC contribution. It might be possible to improve the correction to allow for the expected scaling of the trispectrum, or to dampen it where the trispectrum becomes important. To better recover the low $k$ off-diagonal elements using volume scaling, a higher-order analytic approximation of the form of the covariance matrix could be used to generate the corrective factor. Improving the correction in these ways is left for future work.

\section{Conclusions} \label{Sec_Conclusions}

The super-sample covariance component within covariance matrices is now well understood. Given a set of simulations calculated with fixed background density, it is the missed component of the scatter between simulation results due to background density fluctuations that would exist between patches of the Universe with the same volume and background cosmology as the simulations. Because of the coupling between large and small-scale modes, the effect dominates the contribution to off-diagonal elements of the dark matter power spectrum covariance matrix and substantially correlates on-diagonal elements in the non-linear regime. For galaxy power spectra measured in next-generation surveys, the SSC term is expected to be a significant fraction of the covariance on scales of $k \leq 0.3\ h$Mpc$^{-1}$, beyond which it is largely washed out by shotnoise. Understanding this component is vital when constructing covariance matrices and when considering fast methods for their calculation \cite{Howlett_2017}. 

We have investigated three ways to include SSC within a covariance matrix by either running simulations with different cosmological parameters, or by including an additive term in the estimation of the covariance matrix. Running a set of simulations to include SSC effects is non-trivial \cite{Sirko_2005,Wagner_2015}. The SSC effect can be modeled by changing the cosmological parameters of a simulation in the presence of a background mode. Two methods have been proposed to do this: one based on spherical collapse \cite{Wagner_2015,Zhai_2022}, and one based on a different perturbative expansion linking the background density to cosmological parameters \cite{Sirko_2005}. We have shown that these give very similar results, and match the expected additive contribution to the covariance based on the effect of varying the background on the power spectrum \cite{Li_2014}. The ensemble methods achieve results with fewer total simulations, but with increased complexity of implementation. The additive method offers further flexibility including allowing simpler volume scaling, and should be preferred for any applications where such flexibility is desired. 

Being able to model SSC should, in principle, allow us to scale covariance matrices between samples of different volumes, speeding up the calculation \cite{Howlett_2017}, as this is the component that has a non-trivial scaling with volume. We show that this scaling works using a wider range of simulations than previously considered. In principle, this will allow us to use smaller, less computationally expensive mocks and hence to make higher precision covariance matrices in a given amount of computation time. Using the new simulations, we found that if we push the volume scaling too far, we begin to see effects from the discrete number of modes in small-$k$ bins adjusting the average $k$ value and thus the scale at which $P(k)$ is measured. We discuss this effect and show how we can mitigate it by rescaling $P(k)$ by the $k$-space volume they represent in the covariance. We also see the distribution of power in low $k$ bins becoming significantly non-Gaussian, requiring them to be modeled by a Rayleigh distribution to be usable in computing cosmological parameter estimates. Even so, we are able to achieve a 3\% match on scales $k<1.0\,h^{-1}$Mpc scaling the simulation volume by a factor 512 (see Fig.~\ref{fig_volscaleOndiagLocalByRatio}). These scales include all those of interest for current surveys. The main limitation is the lack of SSC correction for modes lost in the smaller boxes that are present in the large boxes. We note that this level of accuracy is comparable to the agreement between analytic and mock-based covariance estimates to be used with DESI Y1 results \cite{Rashkovetskyi_2024} and thus should be acceptable for future surveys, potentially providing a huge saving in computational resources.

Our analysis has, so far, avoided introducing a number of complications relevant for realistic surveys, particularly window functions and redshift-space effects. Ref.~\cite{Howlett_2017} has shown that the inclusion of these will be non-trivial, and is left for future work. However, we do not expect these to affect the more fundamental volume scaling arguments discussed in our paper. The work here comparing SSC inclusion methods and discussing effects of finite number of modes will remain important.

The simulation parameter files and power spectra used in this work are publicly available at \url{https://zenodo.org/records/14218207}.

\acknowledgments

GS thanks Adrian Bayer and Wayne Hu for helpful discussions. 
AK was supported as a CITA National Fellow by the Natural Sciences and Engineering Research Council of Canada (NSERC), funding reference \#DIS-2022-568580.
SJ acknowledges the Dennis Sciama Fellowship at the University of Portsmouth and the Ram\'{o}n y Cajal Fellowship from the Spanish Ministry of Science.
WP acknowledges the support of the Natural Sciences and Engineering Research Council of Canada (NSERC), [funding reference number RGPIN-2019-03908] and from the Canadian Space Agency.
Research at Perimeter Institute is supported in part by the Government of Canada through the Department of Innovation, Science and Economic Development Canada and by the Province of Ontario through the Ministry of Colleges and Universities.
This research was enabled in part by support provided by Compute Ontario (computeontario.ca) and the Digital Research Alliance of Canada (alliancecan.ca).

\bibliographystyle{JHEP}
\bibliography{references}

\providecommand{\href}[2]{#2}\begingroup\raggedright\begin{thebibliography}{10}

\bibitem{DESI_2016}
{DESI Collaboration}, A.~{Aghamousa}, J.~{Aguilar}, S.~{Ahlen}, S.~{Alam}, L.E.~{Allen} et~al., \emph{{The DESI Experiment Part I: Science,Targeting, and Survey Design}}, \href{https://doi.org/10.48550/arXiv.1611.00036}{\emph{arXiv e-prints} (2016) arXiv:1611.00036} [\href{https://arxiv.org/abs/1611.00036}{{\ttfamily 1611.00036}}].

\bibitem{Euclid_2011}
R.~{Laureijs}, J.~{Amiaux}, S.~{Arduini}, J.L.~{Augu{\`e}res}, J.~{Brinchmann}, R.~{Cole} et~al., \emph{{Euclid Definition Study Report}}, \href{https://doi.org/10.48550/arXiv.1110.3193}{\emph{arXiv e-prints} (2011) arXiv:1110.3193} [\href{https://arxiv.org/abs/1110.3193}{{\ttfamily 1110.3193}}].

\bibitem{LSST}
{\v{Z}}.~{Ivezi{\'c}}, S.M.~{Kahn}, J.A.~{Tyson}, B.~{Abel}, E.~{Acosta}, R.~{Allsman} et~al., \emph{{LSST: From Science Drivers to Reference Design and Anticipated Data Products}}, \href{https://doi.org/10.3847/1538-4357/ab042c}{\emph{\apj} {\bfseries 873} (2019) 111} [\href{https://arxiv.org/abs/0805.2366}{{\ttfamily 0805.2366}}].

\bibitem{roman}
D.~{Spergel}, N.~{Gehrels}, C.~{Baltay}, D.~{Bennett}, J.~{Breckinridge}, M.~{Donahue} et~al., \emph{{Wide-Field InfrarRed Survey Telescope-Astrophysics Focused Telescope Assets WFIRST-AFTA 2015 Report}}, \href{https://doi.org/10.48550/arXiv.1503.03757}{\emph{arXiv e-prints} (2015) arXiv:1503.03757} [\href{https://arxiv.org/abs/1503.03757}{{\ttfamily 1503.03757}}].

\bibitem{spherex}
O.~{Dor{\'e}}, J.~{Bock}, M.~{Ashby}, P.~{Capak}, A.~{Cooray}, R.~{de Putter} et~al., \emph{{Cosmology with the SPHEREX All-Sky Spectral Survey}}, \href{https://doi.org/10.48550/arXiv.1412.4872}{\emph{arXiv e-prints} (2014) arXiv:1412.4872} [\href{https://arxiv.org/abs/1412.4872}{{\ttfamily 1412.4872}}].

\bibitem{BOSS}
K.S.~{Dawson}, D.J.~{Schlegel}, C.P.~{Ahn}, S.F.~{Anderson}, {\'E}.~{Aubourg}, S.~{Bailey} et~al., \emph{{The Baryon Oscillation Spectroscopic Survey of SDSS-III}}, \href{https://doi.org/10.1088/0004-6256/145/1/10}{\emph{\aj} {\bfseries 145} (2013) 10} [\href{https://arxiv.org/abs/1208.0022}{{\ttfamily 1208.0022}}].

\bibitem{White_2015}
M.~{White} and N.~{Padmanabhan}, \emph{{Including parameter dependence in the data and covariance for cosmological inference}}, \href{https://doi.org/10.1088/1475-7516/2015/12/058}{\emph{\jcap} {\bfseries 2015} (2015) 058} [\href{https://arxiv.org/abs/1508.00566}{{\ttfamily 1508.00566}}].

\bibitem{Kalus_2016}
B.~{Kalus}, W.J.~{Percival} and L.~{Samushia}, \emph{{Cosmological parameter inference from galaxy clustering: the effect of the posterior distribution of the power spectrum}}, \href{https://doi.org/10.1093/mnras/stv2307}{\emph{\mnras} {\bfseries 455} (2016) 2573} [\href{https://arxiv.org/abs/1504.03979}{{\ttfamily 1504.03979}}].

\bibitem{Wang2019}
M.S.~{Wang}, W.J.~{Percival}, S.~{Avila}, R.~{Crittenden} and D.~{Bianchi}, \emph{{Cosmological inference from galaxy-clustering power spectrum: Gaussianization and covariance decomposition}}, \href{https://doi.org/10.1093/mnras/stz829}{\emph{\mnras} {\bfseries 486} (2019) 951} [\href{https://arxiv.org/abs/1811.08155}{{\ttfamily 1811.08155}}].

\bibitem{desi-y1-bao}
{DESI Collaboration}, A.G.~{Adame}, J.~{Aguilar}, S.~{Ahlen}, S.~{Alam}, D.M.~{Alexander} et~al., \emph{{DESI 2024 VI: Cosmological Constraints from the Measurements of Baryon Acoustic Oscillations}}, \href{https://doi.org/10.48550/arXiv.2404.03002}{\emph{arXiv e-prints} (2024) arXiv:2404.03002} [\href{https://arxiv.org/abs/2404.03002}{{\ttfamily 2404.03002}}].

\bibitem{FKP1993}
H.A.~{Feldman}, N.~{Kaiser} and J.A.~{Peacock}, \emph{{Power-Spectrum Analysis of Three-dimensional Redshift Surveys}}, \href{https://doi.org/10.1086/174036}{\emph{\apj} {\bfseries 426} (1994) 23} [\href{https://arxiv.org/abs/astro-ph/9304022}{{\ttfamily astro-ph/9304022}}].

\bibitem{Bertolini_2016}
D.~{Bertolini}, K.~{Schutz}, M.P.~{Solon}, J.R.~{Walsh} and K.M.~{Zurek}, \emph{{Non-Gaussian covariance of the matter power spectrum in the effective field theory of large scale structure}}, \href{https://doi.org/10.1103/PhysRevD.93.123505}{\emph{\prd} {\bfseries 93} (2016) 123505} [\href{https://arxiv.org/abs/1512.07630}{{\ttfamily 1512.07630}}].

\bibitem{Carron_2015}
J.~{Carron}, M.~{Wolk} and I.~{Szapudi}, \emph{{On the information content of the matter power spectrum}}, \href{https://doi.org/10.1093/mnras/stv1595}{\emph{\mnras} {\bfseries 453} (2015) 450} [\href{https://arxiv.org/abs/1412.5511}{{\ttfamily 1412.5511}}].

\bibitem{Hou_2022}
J.~{Hou}, R.N.~{Cahn}, O.H.E.~{Philcox} and Z.~{Slepian}, \emph{{Analytic Gaussian covariance matrices for galaxy N -point correlation functions}}, \href{https://doi.org/10.1103/PhysRevD.106.043515}{\emph{\prd} {\bfseries 106} (2022) 043515} [\href{https://arxiv.org/abs/2108.01714}{{\ttfamily 2108.01714}}].

\bibitem{Mohammed_2014}
I.~{Mohammed} and U.~{Seljak}, \emph{{Analytic model for the matter power spectrum, its covariance matrix and baryonic effects}}, \href{https://doi.org/10.1093/mnras/stu1972}{\emph{\mnras} {\bfseries 445} (2014) 3382} [\href{https://arxiv.org/abs/1407.0060}{{\ttfamily 1407.0060}}].

\bibitem{Neyrinck_2011}
M.C.~{Neyrinck}, \emph{{Removable Matter-power-spectrum Covariance from Bias Fluctuations}}, \href{https://doi.org/10.1088/0004-637X/736/1/8}{\emph{\apj} {\bfseries 736} (2011) 8} [\href{https://arxiv.org/abs/1103.5476}{{\ttfamily 1103.5476}}].

\bibitem{Seljak_2000}
U.~{Seljak}, \emph{{Analytic model for galaxy and dark matter clustering}}, \href{https://doi.org/10.1046/j.1365-8711.2000.03715.x}{\emph{\mnras} {\bfseries 318} (2000) 203} [\href{https://arxiv.org/abs/astro-ph/0001493}{{\ttfamily astro-ph/0001493}}].

\bibitem{Taylor_2013}
A.~{Taylor}, B.~{Joachimi} and T.~{Kitching}, \emph{{Putting the precision in precision cosmology: How accurate should your data covariance matrix be?}}, \href{https://doi.org/10.1093/mnras/stt270}{\emph{\mnras} {\bfseries 432} (2013) 1928} [\href{https://arxiv.org/abs/1212.4359}{{\ttfamily 1212.4359}}].

\bibitem{Sugiyama_2020}
N.S.~{Sugiyama}, S.~{Saito}, F.~{Beutler} and H.-J.~{Seo}, \emph{{Perturbation theory approach to predict the covariance matrices of the galaxy power spectrum and bispectrum in redshift space}}, \href{https://doi.org/10.1093/mnras/staa1940}{\emph{\mnras} {\bfseries 497} (2020) 1684} [\href{https://arxiv.org/abs/1908.06234}{{\ttfamily 1908.06234}}].

\bibitem{Wadekar20}
D.~{Wadekar} and R.~{Scoccimarro}, \emph{{Galaxy power spectrum multipoles covariance in perturbation theory}}, \href{https://doi.org/10.1103/PhysRevD.102.123517}{\emph{\prd} {\bfseries 102} (2020) 123517} [\href{https://arxiv.org/abs/1910.02914}{{\ttfamily 1910.02914}}].

\bibitem{Chartier_2021}
N.~{Chartier}, B.~{Wandelt}, Y.~{Akrami} and F.~{Villaescusa-Navarro}, \emph{{CARPool: fast, accurate computation of large-scale structure statistics by pairing costly and cheap cosmological simulations}}, \href{https://doi.org/10.1093/mnras/stab430}{\emph{\mnras} {\bfseries 503} (2021) 1897} [\href{https://arxiv.org/abs/2009.08970}{{\ttfamily 2009.08970}}].

\bibitem{Chartier_2022}
N.~{Chartier} and B.D.~{Wandelt}, \emph{{CARPool covariance: fast, unbiased covariance estimation for large-scale structure observables}}, \href{https://doi.org/10.1093/mnras/stab3097}{\emph{\mnras} {\bfseries 509} (2022) 2220} [\href{https://arxiv.org/abs/2106.11718}{{\ttfamily 2106.11718}}].

\bibitem{Dodelson_2013}
S.~{Dodelson} and M.D.~{Schneider}, \emph{{The effect of covariance estimator error on cosmological parameter constraints}}, \href{https://doi.org/10.1103/PhysRevD.88.063537}{\emph{\prd} {\bfseries 88} (2013) 063537} [\href{https://arxiv.org/abs/1304.2593}{{\ttfamily 1304.2593}}].

\bibitem{Percival_2014}
W.J.~{Percival}, A.J.~{Ross}, A.G.~{S{\'a}nchez}, L.~{Samushia}, A.~{Burden}, R.~{Crittenden} et~al., \emph{{The clustering of Galaxies in the SDSS-III Baryon Oscillation Spectroscopic Survey: including covariance matrix errors}}, \href{https://doi.org/10.1093/mnras/stu112}{\emph{\mnras} {\bfseries 439} (2014) 2531} [\href{https://arxiv.org/abs/1312.4841}{{\ttfamily 1312.4841}}].

\bibitem{Paz_2015}
D.J.~{Paz} and A.G.~{S{\'a}nchez}, \emph{{Improving the precision matrix for precision cosmology}}, \href{https://doi.org/10.1093/mnras/stv2259}{\emph{\mnras} {\bfseries 454} (2015) 4326} [\href{https://arxiv.org/abs/1508.03162}{{\ttfamily 1508.03162}}].

\bibitem{Pope2008}
A.C.~{Pope} and I.~{Szapudi}, \emph{{Shrinkage estimation of the power spectrum covariance matrix}}, \href{https://doi.org/10.1111/j.1365-2966.2008.13561.x}{\emph{\mnras} {\bfseries 389} (2008) 766} [\href{https://arxiv.org/abs/0711.2509}{{\ttfamily 0711.2509}}].

\bibitem{Joachimi2017}
B.~{Joachimi}, \emph{{Non-linear shrinkage estimation of large-scale structure covariance}}, \href{https://doi.org/10.1093/mnrasl/slw240}{\emph{\mnras} {\bfseries 466} (2017) L83} [\href{https://arxiv.org/abs/1612.00752}{{\ttfamily 1612.00752}}].

\bibitem{Gaztanaga2005}
E.~{Gazta{\~n}aga} and R.~{Scoccimarro}, \emph{{The three-point function in large-scale structure: redshift distortions and galaxy bias}}, \href{https://doi.org/10.1111/j.1365-2966.2005.09234.x}{\emph{\mnras} {\bfseries 361} (2005) 824} [\href{https://arxiv.org/abs/astro-ph/0501637}{{\ttfamily astro-ph/0501637}}].

\bibitem{Friedrich2018}
O.~{Friedrich} and T.~{Eifler}, \emph{{Precision matrix expansion - efficient use of numerical simulations in estimating errors on cosmological parameters}}, \href{https://doi.org/10.1093/mnras/stx2566}{\emph{\mnras} {\bfseries 473} (2018) 4150} [\href{https://arxiv.org/abs/1703.07786}{{\ttfamily 1703.07786}}].

\bibitem{Pearson2016}
D.W.~{Pearson} and L.~{Samushia}, \emph{{Estimating the power spectrum covariance matrix with fewer mock samples}}, \href{https://doi.org/10.1093/mnras/stw062}{\emph{\mnras} {\bfseries 457} (2016) 993} [\href{https://arxiv.org/abs/1509.00064}{{\ttfamily 1509.00064}}].

\bibitem{Fumagalli2024}
{Euclid Collaboration}, A.~{Fumagalli}, A.~{Saro}, S.~{Borgani}, T.~{Castro}, M.~{Costanzi} et~al., \emph{{Euclid preparation. XXXV. Covariance model validation for the two-point correlation function of galaxy clusters}}, \href{https://doi.org/10.1051/0004-6361/202245540}{\emph{\aap} {\bfseries 683} (2024) A253} [\href{https://arxiv.org/abs/2211.12965}{{\ttfamily 2211.12965}}].

\bibitem{MOPED}
A.F.~{Heavens}, R.~{Jimenez} and O.~{Lahav}, \emph{{Massive lossless data compression and multiple parameter estimation from galaxy spectra}}, \href{https://doi.org/10.1046/j.1365-8711.2000.03692.x}{\emph{\mnras} {\bfseries 317} (2000) 965} [\href{https://arxiv.org/abs/astro-ph/9911102}{{\ttfamily astro-ph/9911102}}].

\bibitem{Lai-2024}
Y.~{Lai}, C.~{Howlett} and T.M.~{Davis}, \emph{{Faster cosmological analysis with power spectrum without simulations}}, \href{https://doi.org/10.1093/mnras/stae1134}{\emph{\mnras} {\bfseries 530} (2024) 4519} [\href{https://arxiv.org/abs/2306.00388}{{\ttfamily 2306.00388}}].

\bibitem{Howlett_2017}
C.~{Howlett} and W.J.~{Percival}, \emph{{Galaxy two-point covariance matrix estimation for next generation surveys}}, \href{https://doi.org/10.1093/mnras/stx2342}{\emph{\mnras} {\bfseries 472} (2017) 4935} [\href{https://arxiv.org/abs/1709.03057}{{\ttfamily 1709.03057}}].

\bibitem{Klypin_2018}
A.~{Klypin} and F.~{Prada}, \emph{{Dark matter statistics for large galaxy catalogues: power spectra and covariance matrices}}, \href{https://doi.org/10.1093/mnras/sty1340}{\emph{\mnras} {\bfseries 478} (2018) 4602} [\href{https://arxiv.org/abs/1701.05690}{{\ttfamily 1701.05690}}].

\bibitem{Klypin_2019}
A.~{Klypin} and F.~{Prada}, \emph{{Effects of long-wavelength fluctuations in large galaxy surveys}}, \href{https://doi.org/10.1093/mnras/stz2194}{\emph{\mnras} {\bfseries 489} (2019) 1684} [\href{https://arxiv.org/abs/1809.03637}{{\ttfamily 1809.03637}}].

\bibitem{DongPaez24}
C.A.~{Dong-P{\'a}ez}, A.~{Smith}, A.O.~{Szewciw}, J.~{Ereza}, M.H.~{Abdullah}, C.~{Hern{\'a}ndez-Aguayo} et~al., \emph{{The Uchuu-SDSS galaxy light-cones: a clustering, redshift space distortion and baryonic acoustic oscillation study}}, \href{https://doi.org/10.1093/mnras/stae062}{\emph{\mnras} {\bfseries 528} (2024) 7236} [\href{https://arxiv.org/abs/2208.00540}{{\ttfamily 2208.00540}}].

\bibitem{Ereza_2024}
J.~{Ereza}, F.~{Prada}, A.~{Klypin}, T.~{Ishiyama}, A.~{Smith}, C.M.~{Baugh} et~al., \emph{{The UCHUU-GLAM BOSS and eBOSS LRG lightcones: exploring clustering and covariance errors}}, \href{https://doi.org/10.1093/mnras/stae1543}{\emph{\mnras} {\bfseries 532} (2024) 1659} [\href{https://arxiv.org/abs/2311.14456}{{\ttfamily 2311.14456}}].

\bibitem{Meiksin_1999}
A.~{Meiksin}, M.~{White} and J.A.~{Peacock}, \emph{{Baryonic signatures in large-scale structure}}, \href{https://doi.org/10.1046/j.1365-8711.1999.02369.x}{\emph{\mnras} {\bfseries 304} (1999) 851} [\href{https://arxiv.org/abs/astro-ph/9812214}{{\ttfamily astro-ph/9812214}}].

\bibitem{Scoccimarro_1999}
R.~{Scoccimarro}, M.~{Zaldarriaga} and L.~{Hui}, \emph{{Power Spectrum Correlations Induced by Nonlinear Clustering}}, \href{https://doi.org/10.1086/308059}{\emph{\apj} {\bfseries 527} (1999) 1} [\href{https://arxiv.org/abs/astro-ph/9901099}{{\ttfamily astro-ph/9901099}}].

\bibitem{Takada_2013}
M.~Takada and W.~Hu, \emph{Power spectrum super-sample covariance}, \href{https://doi.org/10.1103/PhysRevD.87.123504}{\emph{Phys. Rev. D} {\bfseries 87} (2013) 123504}.

\bibitem{Hu_2003}
W.~{Hu} and A.V.~{Kravtsov}, \emph{{Sample Variance Considerations for Cluster Surveys}}, \href{https://doi.org/10.1086/345846}{\emph{\apj} {\bfseries 584} (2003) 702} [\href{https://arxiv.org/abs/astro-ph/0203169}{{\ttfamily astro-ph/0203169}}].

\bibitem{Hamilton_2006}
A.J.S.~{Hamilton}, C.D.~{Rimes} and R.~{Scoccimarro}, \emph{{On measuring the covariance matrix of the non-linear power spectrum from simulations}}, \href{https://doi.org/10.1111/j.1365-2966.2006.10709.x}{\emph{\mnras} {\bfseries 371} (2006) 1188} [\href{https://arxiv.org/abs/astro-ph/0511416}{{\ttfamily astro-ph/0511416}}].

\bibitem{Gnedin_2011}
N.Y.~{Gnedin}, A.V.~{Kravtsov} and D.H.~{Rudd}, \emph{{Implementing the DC Mode in Cosmological Simulations with Supercomoving Variables}}, \href{https://doi.org/10.1088/0067-0049/194/2/46}{\emph{\apjs} {\bfseries 194} (2011) 46} [\href{https://arxiv.org/abs/1104.1428}{{\ttfamily 1104.1428}}].

\bibitem{Takahashi_2009}
R.~{Takahashi}, N.~{Yoshida}, M.~{Takada}, T.~{Matsubara}, N.~{Sugiyama}, I.~{Kayo} et~al., \emph{{Simulations of Baryon Acoustic Oscillations. II. Covariance Matrix of the Matter Power Spectrum}}, \href{https://doi.org/10.1088/0004-637X/700/1/479}{\emph{\apj} {\bfseries 700} (2009) 479} [\href{https://arxiv.org/abs/0902.0371}{{\ttfamily 0902.0371}}].

\bibitem{Baldauf_2011}
T.~{Baldauf}, U.~{Seljak}, L.~{Senatore} and M.~{Zaldarriaga}, \emph{{Galaxy bias and non-linear structure formation in general relativity}}, \href{https://doi.org/10.1088/1475-7516/2011/10/031}{\emph{\jcap} {\bfseries 2011} (2011) 031} [\href{https://arxiv.org/abs/1106.5507}{{\ttfamily 1106.5507}}].

\bibitem{Li_2014}
Y.~Li, W.~Hu and M.~Takada, \emph{Super-sample covariance in simulations}, \href{https://doi.org/10.1103/PhysRevD.89.083519}{\emph{Phys. Rev. D} {\bfseries 89} (2014) 083519}.

\bibitem{Wagner_2015}
C.~{Wagner}, F.~{Schmidt}, C.T.~{Chiang} and E.~{Komatsu}, \emph{{Separate universe simulations.}}, \href{https://doi.org/10.1093/mnrasl/slu187}{\emph{\mnras} {\bfseries 448} (2015) L11} [\href{https://arxiv.org/abs/1409.6294}{{\ttfamily 1409.6294}}].

\bibitem{Akitsu_2017}
K.~{Akitsu}, M.~{Takada} and Y.~{Li}, \emph{{Large-scale tidal effect on redshift-space power spectrum in a finite-volume survey}}, \href{https://doi.org/10.1103/PhysRevD.95.083522}{\emph{\prd} {\bfseries 95} (2017) 083522} [\href{https://arxiv.org/abs/1611.04723}{{\ttfamily 1611.04723}}].

\bibitem{Akitsu_2018}
K.~{Akitsu} and M.~{Takada}, \emph{{Impact of large-scale tides on cosmological distortions via redshift-space power spectrum}}, \href{https://doi.org/10.1103/PhysRevD.97.063527}{\emph{\prd} {\bfseries 97} (2018) 063527} [\href{https://arxiv.org/abs/1711.00012}{{\ttfamily 1711.00012}}].

\bibitem{Barreira_2017a}
A.~{Barreira} and F.~{Schmidt}, \emph{{Responses in large-scale structure}}, \href{https://doi.org/10.1088/1475-7516/2017/06/053}{\emph{\jcap} {\bfseries 2017} (2017) 053} [\href{https://arxiv.org/abs/1703.09212}{{\ttfamily 1703.09212}}].

\bibitem{Barreira_2017b}
A.~{Barreira} and F.~{Schmidt}, \emph{{Response approach to the matter power spectrum covariance}}, \href{https://doi.org/10.1088/1475-7516/2017/11/051}{\emph{\jcap} {\bfseries 2017} (2017) 051} [\href{https://arxiv.org/abs/1705.01092}{{\ttfamily 1705.01092}}].

\bibitem{Chan_2018}
K.C.~{Chan}, A.~{Moradinezhad Dizgah} and J.~{Nore{\~n}a}, \emph{{Bispectrum supersample covariance}}, \href{https://doi.org/10.1103/PhysRevD.97.043532}{\emph{\prd} {\bfseries 97} (2018) 043532} [\href{https://arxiv.org/abs/1709.02473}{{\ttfamily 1709.02473}}].

\bibitem{Li_2018}
Y.~{Li}, M.~{Schmittfull} and U.~{Seljak}, \emph{{Galaxy power-spectrum responses and redshift-space super-sample effect}}, \href{https://doi.org/10.1088/1475-7516/2018/02/022}{\emph{\jcap} {\bfseries 2018} (2018) 022} [\href{https://arxiv.org/abs/1711.00018}{{\ttfamily 1711.00018}}].

\bibitem{Barreira_2018}
A.~{Barreira}, E.~{Krause} and F.~{Schmidt}, \emph{{Complete super-sample lensing covariance in the response approach}}, \href{https://doi.org/10.1088/1475-7516/2018/06/015}{\emph{\jcap} {\bfseries 2018} (2018) 015} [\href{https://arxiv.org/abs/1711.07467}{{\ttfamily 1711.07467}}].

\bibitem{Barreira_2019a}
A.~{Barreira}, \emph{{The squeezed matter bispectrum covariance with responses}}, \href{https://doi.org/10.1088/1475-7516/2019/03/008}{\emph{\jcap} {\bfseries 2019} (2019) 008} [\href{https://arxiv.org/abs/1901.01243}{{\ttfamily 1901.01243}}].

\bibitem{Barreira_2019b}
A.~{Barreira}, D.~{Nelson}, A.~{Pillepich}, V.~{Springel}, F.~{Schmidt}, R.~{Pakmor} et~al., \emph{{Separate Universe simulations with IllustrisTNG: baryonic effects on power spectrum responses and higher-order statistics}}, \href{https://doi.org/10.1093/mnras/stz1807}{\emph{\mnras} {\bfseries 488} (2019) 2079} [\href{https://arxiv.org/abs/1904.02070}{{\ttfamily 1904.02070}}].

\bibitem{Lacasa_2019}
F.~{Lacasa} and J.~{Grain}, \emph{{Fast and easy super-sample covariance of large-scale structure observables}}, \href{https://doi.org/10.1051/0004-6361/201834343}{\emph{\aap} {\bfseries 624} (2019) A61} [\href{https://arxiv.org/abs/1809.05437}{{\ttfamily 1809.05437}}].

\bibitem{Castorina_2020}
E.~{Castorina} and A.~{Moradinezhad Dizgah}, \emph{{Local Primordial Non-Gaussianities and super-sample variance}}, \href{https://doi.org/10.1088/1475-7516/2020/10/007}{\emph{\jcap} {\bfseries 2020} (2020) 007} [\href{https://arxiv.org/abs/2005.14677}{{\ttfamily 2005.14677}}].

\bibitem{Philcox_2020}
O.H.E.~{Philcox}, D.N.~{Spergel} and F.~{Villaescusa-Navarro}, \emph{{Effective halo model: Creating a physical and accurate model of the matter power spectrum and cluster counts}}, \href{https://doi.org/10.1103/PhysRevD.101.123520}{\emph{\prd} {\bfseries 101} (2020) 123520} [\href{https://arxiv.org/abs/2004.09515}{{\ttfamily 2004.09515}}].

\bibitem{Halder_2022}
A.~{Halder} and A.~{Barreira}, \emph{{Response approach to the integrated shear 3-point correlation function: the impact of baryonic effects on small scales}}, \href{https://doi.org/10.1093/mnras/stac2046}{\emph{\mnras} {\bfseries 515} (2022) 4639} [\href{https://arxiv.org/abs/2201.05607}{{\ttfamily 2201.05607}}].

\bibitem{Zhai_2022}
Z.~Zhai and W.J.~Percival, \emph{Sample variance for supernovae distance measurements and the hubble tension}, \href{https://doi.org/10.1103/physrevd.106.103527}{\emph{Physical Review D} {\bfseries 106} (2022) }.

\bibitem{Bayer_2022}
A.E.~{Bayer}, J.~{Liu}, R.~{Terasawa}, A.~{Barreira}, Y.~{Zhong} and Y.~{Feng}, \emph{{Super-sample covariance of the power spectrum, bispectrum, halos, voids, and their cross covariances}}, \href{https://doi.org/10.1103/PhysRevD.108.043521}{\emph{\prd} {\bfseries 108} (2023) 043521} [\href{https://arxiv.org/abs/2210.15647}{{\ttfamily 2210.15647}}].

\bibitem{Gouyou_2022}
S.~{Gouyou Beauchamps}, F.~{Lacasa}, I.~{Tutusaus}, M.~{Aubert}, P.~{Baratta}, A.~{Gorce} et~al., \emph{{Impact of survey geometry and super-sample covariance on future photometric galaxy surveys}}, \href{https://doi.org/10.1051/0004-6361/202142052}{\emph{\aap} {\bfseries 659} (2022) A128} [\href{https://arxiv.org/abs/2109.02308}{{\ttfamily 2109.02308}}].

\bibitem{Terasawa_2022}
R.~{Terasawa}, R.~{Takahashi}, T.~{Nishimichi} and M.~{Takada}, \emph{{Separate universe approach to evaluate nonlinear matter power spectrum for nonflat {\ensuremath{\Lambda}} CDM model}}, \href{https://doi.org/10.1103/PhysRevD.106.083504}{\emph{\prd} {\bfseries 106} (2022) 083504} [\href{https://arxiv.org/abs/2205.10339}{{\ttfamily 2205.10339}}].

\bibitem{Linke_2024}
L.~{Linke}, P.A.~{Burger}, S.~{Heydenreich}, L.~{Porth} and P.~{Schneider}, \emph{{What is the super-sample covariance? A fresh perspective for second-order shear statistics}}, \href{https://doi.org/10.1051/0004-6361/202346225}{\emph{\aap} {\bfseries 681} (2024) A33} [\href{https://arxiv.org/abs/2302.12277}{{\ttfamily 2302.12277}}].

\bibitem{Desjacques_2018}
V.~{Desjacques}, D.~{Jeong} and F.~{Schmidt}, \emph{{Large-scale galaxy bias}}, \href{https://doi.org/10.1016/j.physrep.2017.12.002}{\emph{\physrep} {\bfseries 733} (2018) 1} [\href{https://arxiv.org/abs/1611.09787}{{\ttfamily 1611.09787}}].

\bibitem{Sirko_2005}
E.~Sirko, \emph{Initial conditions to cosmological $\emph{N}$-body simulations, or, how to run an ensemble of simulations}, \href{https://doi.org/10.1086/497090}{\emph{The Astrophysical Journal} {\bfseries 634} (2005) 728}.

\bibitem{Mcdonald_2003}
P.~{McDonald}, \emph{{Toward a Measurement of the Cosmological Geometry at z \raisebox{-0.5ex}\textasciitilde 2: Predicting Ly{\ensuremath{\alpha}} Forest Correlation in Three Dimensions and the Potential of Future Data Sets}}, \href{https://doi.org/10.1086/345945}{\emph{\apj} {\bfseries 585} (2003) 34} [\href{https://arxiv.org/abs/astro-ph/0108064}{{\ttfamily astro-ph/0108064}}].

\bibitem{Goldberg_2004}
D.M.~{Goldberg} and M.S.~{Vogeley}, \emph{{Simulating Voids}}, \href{https://doi.org/10.1086/382143}{\emph{\apj} {\bfseries 605} (2004) 1} [\href{https://arxiv.org/abs/astro-ph/0307191}{{\ttfamily astro-ph/0307191}}].

\bibitem{Martino_2009}
M.C.~{Martino} and R.K.~{Sheth}, \emph{{On the equivalence between the effective cosmology and excursion set treatments of environment}}, \href{https://doi.org/10.1111/j.1365-2966.2009.14467.x}{\emph{\mnras} {\bfseries 394} (2009) 2109} [\href{https://arxiv.org/abs/0901.0757}{{\ttfamily 0901.0757}}].

\bibitem{Dai_2015}
L.~{Dai}, E.~{Pajer} and F.~{Schmidt}, \emph{{On separate universes}}, \href{https://doi.org/10.1088/1475-7516/2015/10/059}{\emph{\jcap} {\bfseries 2015} (2015) 059} [\href{https://arxiv.org/abs/1504.00351}{{\ttfamily 1504.00351}}].

\bibitem{Tormen_1996}
G.~{Tormen} and E.~{Bertschinger}, \emph{{Adding Long-Wavelength Modes to an N-Body Simulation}}, \href{https://doi.org/10.1086/178037}{\emph{\apj} {\bfseries 472} (1996) 14} [\href{https://arxiv.org/abs/astro-ph/9512131}{{\ttfamily astro-ph/9512131}}].

\bibitem{Cole_1997}
S.~{Cole}, \emph{{Adding Long-Wavelength Power to N-body simulations}}, \href{https://doi.org/10.1093/mnras/286.1.38}{\emph{\mnras} {\bfseries 286} (1997) 38} [\href{https://arxiv.org/abs/astro-ph/9604046}{{\ttfamily astro-ph/9604046}}].

\bibitem{Percival_2005}
W.J.~{Percival}, \emph{{Cosmological structure formation in a homogeneous dark energy background}}, \href{https://doi.org/10.1051/0004-6361:20053637}{\emph{\aap} {\bfseries 443} (2005) 819} [\href{https://arxiv.org/abs/astro-ph/0508156}{{\ttfamily astro-ph/0508156}}].

\bibitem{LPICOLA}
C.~{Howlett}, M.~{Manera} and W.J.~{Percival}, \emph{{L-PICOLA: A parallel code for fast dark matter simulation}}, \href{https://doi.org/10.1016/j.ascom.2015.07.003}{\emph{Astronomy and Computing} {\bfseries 12} (2015) 109} [\href{https://arxiv.org/abs/1506.03737}{{\ttfamily 1506.03737}}].

\bibitem{Tassev_2013}
S.~{Tassev}, M.~{Zaldarriaga} and D.J.~{Eisenstein}, \emph{{Solving large scale structure in ten easy steps with COLA}}, \href{https://doi.org/10.1088/1475-7516/2013/06/036}{\emph{\jcap} {\bfseries 2013} (2013) 036} [\href{https://arxiv.org/abs/1301.0322}{{\ttfamily 1301.0322}}].

\bibitem{Hand2018_nbodykit}
N.~Hand, Y.~Feng, F.~Beutler, Y.~Li, C.~Modi, U.~Seljak et~al., \emph{nbodykit: An open-source, massively parallel toolkit for large-scale structure}, {\emph{The Astronomical Journal} {\bfseries 156} (2018) 160}.

\bibitem{dePutter_2012}
R.~{de Putter}, C.~{Wagner}, O.~{Mena}, L.~{Verde} and W.J.~{Percival}, \emph{{Thinking outside the box: effects of modes larger than the survey on matter power spectrum covariance}}, \href{https://doi.org/10.1088/1475-7516/2012/04/019}{\emph{\jcap} {\bfseries 2012} (2012) 019} [\href{https://arxiv.org/abs/1111.6596}{{\ttfamily 1111.6596}}].

\bibitem{Hu_2001}
W.~{Hu} and M.~{White}, \emph{{Power Spectra Estimation for Weak Lensing}}, \href{https://doi.org/10.1086/321380}{\emph{\apj} {\bfseries 554} (2001) 67} [\href{https://arxiv.org/abs/astro-ph/0010352}{{\ttfamily astro-ph/0010352}}].

\bibitem{Springel_2005}
V.~{Springel}, \emph{{The cosmological simulation code GADGET-2}}, \href{https://doi.org/10.1111/j.1365-2966.2005.09655.x}{\emph{\mnras} {\bfseries 364} (2005) 1105} [\href{https://arxiv.org/abs/astro-ph/0505010}{{\ttfamily astro-ph/0505010}}].

\bibitem{Takahashi_2011}
R.~{Takahashi}, N.~{Yoshida}, M.~{Takada}, T.~{Matsubara}, N.~{Sugiyama}, I.~{Kayo} et~al., \emph{{Non-Gaussian Error Contribution to Likelihood Analysis of the Matter Power Spectrum}}, \href{https://doi.org/10.1088/0004-637X/726/1/7}{\emph{\apj} {\bfseries 726} (2011) 7} [\href{https://arxiv.org/abs/0912.1381}{{\ttfamily 0912.1381}}].

\bibitem{Rashkovetskyi_2024}
M.~{Rashkovetskyi}, D.~{Forero-S{\'a}nchez}, A.~{de Mattia}, D.J.~{Eisenstein}, N.~{Padmanabhan}, H.~{Seo} et~al., \emph{{Semi-analytical covariance matrices for two-point correlation function for DESI 2024 data}}, \href{https://doi.org/10.48550/arXiv.2404.03007}{\emph{arXiv e-prints} (2024) arXiv:2404.03007} [\href{https://arxiv.org/abs/2404.03007}{{\ttfamily 2404.03007}}].

\bibitem{Bond_2000}
J.R.~{Bond}, A.H.~{Jaffe} and L.~{Knox}, \emph{{Radical Compression of Cosmic Microwave Background Data}}, \href{https://doi.org/10.1086/308625}{\emph{\apj} {\bfseries 533} (2000) 19} [\href{https://arxiv.org/abs/astro-ph/9808264}{{\ttfamily astro-ph/9808264}}].

\bibitem{Smith_2006}
S.~{Smith}, A.~{Challinor} and G.~{Rocha}, \emph{{What can be learned from the lensed cosmic microwave background B-mode polarization power spectrum?}}, \href{https://doi.org/10.1103/PhysRevD.73.023517}{\emph{\prd} {\bfseries 73} (2006) 023517} [\href{https://arxiv.org/abs/astro-ph/0511703}{{\ttfamily astro-ph/0511703}}].

\bibitem{Percival_2006}
W.J.~{Percival} and M.L.~{Brown}, \emph{{Likelihood techniques for the combined analysis of CMB temperature and polarization power spectra}}, \href{https://doi.org/10.1111/j.1365-2966.2006.10910.x}{\emph{\mnras} {\bfseries 372} (2006) 1104} [\href{https://arxiv.org/abs/astro-ph/0604547}{{\ttfamily astro-ph/0604547}}].

\bibitem{Hamimeche_2008}
S.~{Hamimeche} and A.~{Lewis}, \emph{{Likelihood analysis of CMB temperature and polarization power spectra}}, \href{https://doi.org/10.1103/PhysRevD.77.103013}{\emph{\prd} {\bfseries 77} (2008) 103013} [\href{https://arxiv.org/abs/0801.0554}{{\ttfamily 0801.0554}}].

\bibitem{Kalus_2023}
B.~{Bahr-Kalus}, D.~{Parkinson} and E.-M.~{Mueller}, \emph{{Measurement of the matter-radiation equality scale using the extended baryon oscillation spectroscopic survey quasar sample}}, \href{https://doi.org/10.1093/mnras/stad1867}{\emph{\mnras} {\bfseries 524} (2023) 2463} [\href{https://arxiv.org/abs/2302.07484}{{\ttfamily 2302.07484}}].

\end{thebibliography}\endgroup

\end{document}